\documentclass[universe,article,accept,pdftex,moreauthors]{Definitions/mdpi}

\firstpage{1} 
\pubvolume{10}
\issuenum{11}
\articlenumber{431}
\pubyear{2024}
\copyrightyear{2024}
\externaleditor{Academic Editor: Stephen J. Curran}
\datereceived{14 October 2024} 
\daterevised{16 November 2024} 
\dateaccepted{18 November 2024} 
\datepublished{19 November 2024} 
\hreflink{https://doi.org/10.3390/\linebreak universe10110431} 

\usepackage{amsmath} 
\newcommand{\angstrom}{{\rm \mathring A}}
\usepackage{aas_macros}

\newcommand{\HI}{\hbox{H{\sc i}}}
\newcommand{\Lya}{\hbox{Ly$\alpha$}}
\newcommand{\Ha}{\hbox{H$\alpha$}}
\newcommand{\Hb}{\hbox{H$\beta$}}

\newcommand{\MgIItsne}{\hbox{Mg{\sc ii}}~2798}
\newcommand{\CIV}{\hbox{C{\sc iv}}}
\newcommand{\CIVoffn}{\hbox{C{\sc iv}}~1549}
\newcommand{\HeI}{\hbox{He{\sc i}}}
\newcommand{\HeII}{\hbox{He{\sc ii}}}
\newcommand{\HeIIosfz}{\hbox{He{\sc ii}}~1640}

\Title{The Composite Spectral Energy Distribution of Quasars is Surprisingly Universal Since Cosmic Noon}

\TitleCitation{The Composite Spectral Energy Distribution of Quasars is Surprisingly Universal Since Cosmic Noon}

\Author{{{Zhenyi Cai}} 
 $^{1,2}$\orcidA{}}

\AuthorNames{Zhenyi Cai}

\AuthorCitation{{Cai,} 
 Z.}

\address{%
$^{1}$ \quad Department of Astronomy, University of Science and Technology of China, Hefei 230026, China; zcai@ustc.edu.cn
\\
$^{2}$ \quad School of Astronomy and Space Science, University of Science and Technology of China, Hefei 230026, China
}

\abstract{
{Leveraging the} photometric data of the Sloan Digital Sky Survey and the Galaxy Evolution Explorer (GALEX), {we construct mean/median spectral energy distributions (SEDs) for unique bright quasars in redshift bins of 0.2 and up to $z \simeq 3$}, after taking the GALEX non-detection into account. Further correcting for the absorption of the intergalactic medium, these mean/median quasar SEDs constitute a surprisingly redshift-independent mean/median composite SED from the rest-frame optical down to $\simeq$$500 \angstrom$ for quasars with bolometric luminosity brighter than $10^{45.5} {\rm erg s^{-1}}$. Moreover, the mean/median composite quasar SED is plausibly also independent of black hole mass and Eddington ratio, and suggests similar properties of dust and gas in the quasar host galaxies since cosmic noon. Both the mean and median composite SEDs are nicely consistent with previous mean composite quasar spectra at wavelengths beyond $\simeq$$1000 \angstrom$, but at shorter wavelengths, are redder, indicating, on average, less ionizing radiation than previously expected. {Through comparing the model-predicted to the observed composite quasar SEDs, we favor a simply truncated disk model, rather than a standard thin disk model, for the quasar central engine, though we request more sophisticated disk models. Future deep ultraviolet facilities, such as the China Space Station Telescope and the Ultraviolet Explorer, would prompt revolutions in many aspects, including the quasar central engine, production of the broad emission lines in quasars, and cosmic reionization.}
}


\begin{document}


\section{Introduction} \label{sect:intro}

A quasar, or luminous active galactic nucleus (AGN), is a supermassive black hole (BH) actively accreting in a tiny region of a galactic center. The enormous and long-lasting luminosity radiated by a quasar makes it outshine its host galaxy and become the brightest persistent object glowing in the universe. No sooner had quasars been discovered about 60 years ago \cite{Schmidt1963Natur.197.1040S,Matthews1963ApJ...138...30M} than the basic central engine for their powerful emissions was proposed \cite{Salpeter1964ApJ...140..796S,Lynden-Bell1969Natur.223..690L,Rees1984ARA&A..22..471R}. {{The} 
 quasar central engine} consists of a supermassive BH surrounded by an accretion flow/disk. 

While the exact details of the accretion flow are still being studied, the major part of the accretion disk is likely cold and optically thick \cite{Shakura1973A&A....24..337S,Novikov1973blho.conf..343N}, producing the thermal continuum emission from the rest-frame optical to ultraviolet (UV) wavelengths \cite{Shields1978Natur.272..706S,Malkan1982ApJ...254...22M,Elvis1994ApJS...95....1E}, whereas the innermost region of the accretion flow could be an optically thin hot corona \cite{Galeev1979ApJ...229..318G,Haardt1991ApJ...380L..51H,Haardt1993ApJ...413..507H}, accounting for the non-thermal X-ray emission of quasars \cite{Walter1993A&A...274..105W,Lusso2016ApJ...819..154L,Kang2022ApJ...929..141K}. 
Meanwhile, a warm corona between the cold disk and the hot corona (or atop the cold disk) could be responsible for the soft X-ray excess (\cite{Magdziarz1998MNRAS.301..179M,Done2012MNRAS.420.1848D,Jin2012MNRAS.420.1825J,Petrucci2013A&A...549A..73P,Kubota2018MNRAS.480.1247K}; however, see \cite{Gierlinski2004MNRAS.349L...7G,Crummy2006MNRAS.365.1067C} for the absorption and reflection explanations, respectively). 
Up to now, investigating the multi-wavelength spectral energy distribution (SED) of quasars has been one of the important avenues for uncovering the disk-corona coupling mystery of their central engine \cite{Liu2022iSci...25j3544L}.

\begin{figure}[H]
    \includegraphics[width=0.95\linewidth]{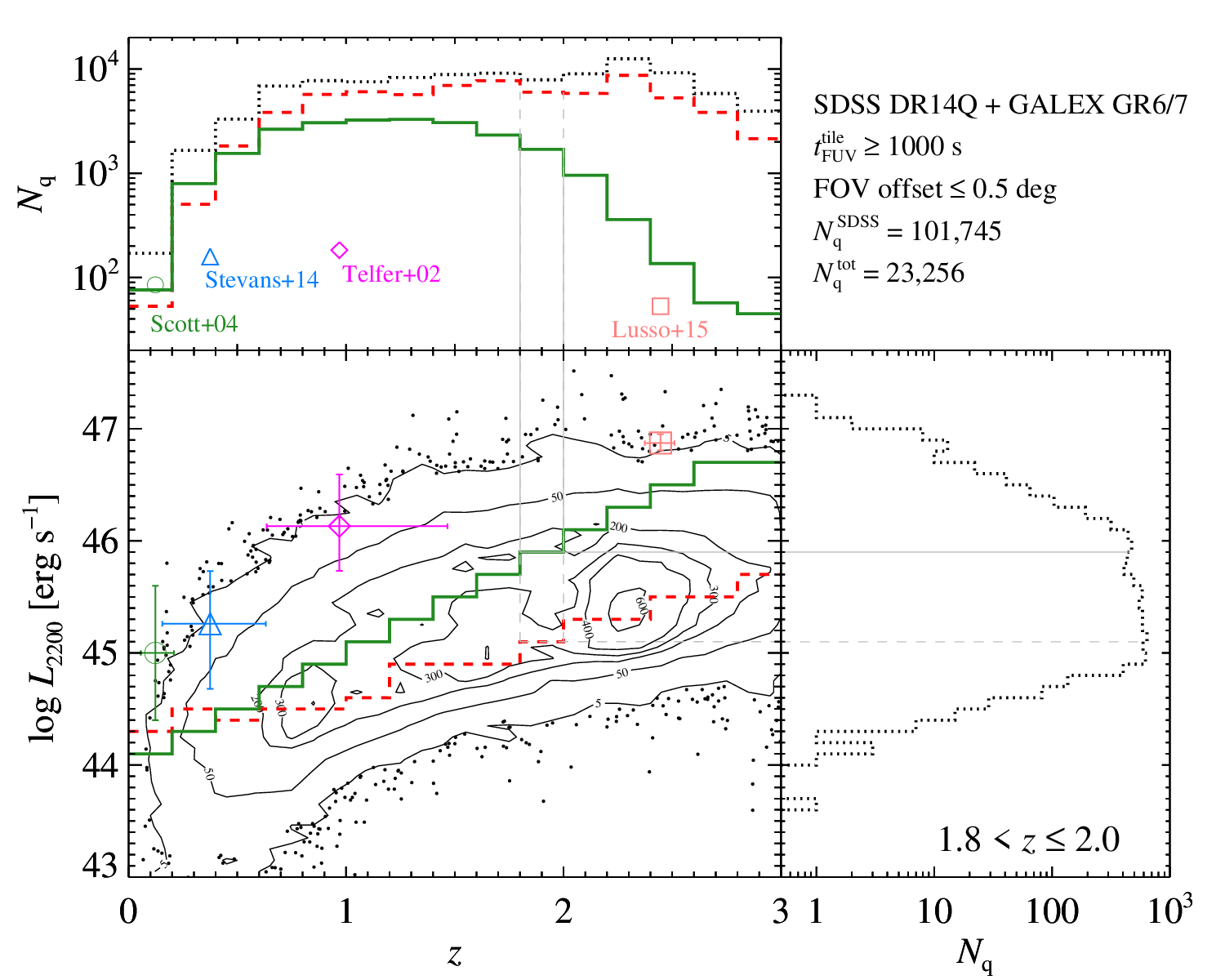}
    \caption{{The} 
 (\textbf{main}) panel presents distributions of our parent quasars (101,745; contours surrounded by sparse dots) and unique bright quasars (23,256; brighter than the green solid stepwise curve) in the luminosity-redshift space, while the (\textbf{top}) panel shows their redshift distributions, i.e., the black dotted histogram and the green solid histogram, respectively. The red dashed stepwise curve in the (\textbf{main}) panel indicates the peak of the luminosity distribution of quasars as a function of redshift, e.g., the peak of the dotted histogram for quasars in $1.8 < z \leq 2.0$ in the (\textbf{right}) panel. The red dashed histogram in the (\textbf{top}) panel is the redshift distribution of quasars brighter than the red dashed stepwise curve.
    For comparison, the (\textbf{main}) panel includes the medians and 25--75th percentile ranges of redshift and UV luminosity of four representative quasar samples (i.e., a circle for \citet{Scott2004ApJ}, a triangle for \citet{Stevans2014ApJ}, a diamond for \citet{Telfer2002ApJ}, and a square for \citet{Lusso2015MNRAS}), while the (\textbf{top}) panel contains the corresponding quasar numbers.}
    \label{fig:quasar_selection}
\end{figure}

In terms of the available spectroscopic or photometric data, there are many composite spectra \cite{Cristiani1990A&A...227..385C,Boyle1990MNRAS.243..231B,Francis1991ApJ...373..465F,Zheng1997ApJ...475..469Z,VandenBerk2001AJ....122..549V,Brotherton2001ApJ...546..775B,Telfer2002ApJ,Scott2004ApJ,Glikman2006ApJ...640..579G,Shull2012ApJ...752..162S,Stevans2014ApJ,Lusso2015MNRAS,Selsing2016A&A...585A..87S} or broadband SEDs \cite{Richards2006ApJS..166..470R,Trammell2007AJ....133.1780T,Krawczyk2013ApJS..206....4K} for quasars, respectively. {The average radiative properties of quasars have been extensively explored in} a broad wavelength range spanning from the near-infrared to X-ray, except the extreme UV (EUV; the rest-frame $\sim$$1000 \angstrom$--$100 \angstrom$). {This is because EUV} observation is very difficult owing to the significant absorption of the neutral hydrogen in our Milky Way as well as in the intergalactic medium (IGM) along the line of sight (LOS) to quasars. 
To circumvent the problem of the hydrogen absorption, {one needs} the rest-frame EUV-involved spectra of individual quasars (only $\sim$$500$ from local to {$z \sim 2.5$} 
 according to \cite{Telfer2002ApJ,Scott2004ApJ,Stevans2014ApJ,Lusso2015MNRAS}; see also Figure \ref{fig:quasar_selection}) {in order to} carefully correct for the IGM absorption before building composite spectra. {It has long been suggested that the EUV-involved composite spectra of quasars with different UV luminosity} are rather similar in shape at wavelengths beyond $\simeq$$1000 \angstrom$ (optical-to-FUV), but exhibit a luminosity dependence in the EUV: the EUV shape {is on average} redder/softer for more luminous quasars. {Later broadband analyses based on the photometric data of the Sloan Digital Sky Survey (SDSS) and the Galaxy Evolution Explorer (GALEX) confirmed the luminosity dependence in the EUV \cite{Trammell2007AJ....133.1780T,Krawczyk2013ApJS..206....4K}.}
\vspace{-6pt}

However, both previous spectral and photometric analyses of the quasar EUV SEDs only consider quasars detected in the EUV and thus are subject to a prominent selection bias: the more luminous quasars with redder EUV shapes are more likely detected. Consequently, the reported luminosity dependence of the quasar EUV SEDs could be naturally expected given such a bias. {Actually, \citet{VandenBerk2020MNRAS.493.2745V} have demonstrated that the EUV color of quasars is substantially redder than expected once taking this bias into account}.

Recently, after controlling for this bias, \citet{Cai2023NatAs} unveiled that the average quasar SEDs down to the rest-frame $\simeq$$500 \angstrom$ are nearly luminosity-independent for $z$$\sim$$2$ quasars with the rest-frame $2200 \angstrom$ monochromatic luminosity brighter than $10^{45} {\rm erg s^{-1}}$, i.e., $\log L_{2200} \geqslant 45$, where $L_{2200} \equiv \nu L_\nu(\lambda_{\rm rest}=2200 \angstrom)$. Furthermore, after considering the EUV non-detection and correcting for the IGM absorption for a unique bright sub-sample of $z$$\sim$$2$ quasars, i.e., with $\log L_{2200} \geqslant 46$, {the unveiled intrinsic average quasar SED is} redder in the EUV than previous quasar composite spectra. The new intrinsic average quasar SED does not match predictions of the standard thin disk model {in which the inner disk boundary extends} to the innermost stable circular orbit. Instead, it prefers a simply truncated disk model, such as the windy disk proposed by \citet{Laor2014MNRAS.438.3024L}, where a microscopic atomic origin is responsible for the disk truncation such that the resultant SEDs are weakly dependent on the quasar physical properties, e.g., the BH mass, $M_{\rm BH}$, and the Eddington ratio, $\lambda_{\rm Edd}$.

Here, we extend our previous analysis of $z$$\sim$$2$ quasars to quasars at $z \leqslant 3$ or since cosmic noon. Utilizing the photometric data of the 14th release of the SDSS quasar catalog (DR14Q; \cite{Paris2018A&A...613A..51P}) and the associated GALEX legacy data \cite{Martin2005ApJ...619L...1M}, Section \ref{sect:methods} presents our selection criteria for the unique bright quasars in different redshift bins and the resultant exciting luminosity- and redshift-independent composite quasar SED after considering the GALEX non-detection and correcting for the IGM absorption. Discussion and implications on the universal composite quasar SED are given in Section \ref{sect:discussion}, {and} Section \ref{sect:conclusion} briefly summarizes our conclusions.
Throughout, we assume a flat cosmology with Hubble constant $H_0 = 70 {\rm km s^{-1} Mpc^{-1}}$ and dark energy density $\Omega_\Lambda = 0.7$.

\section{Quasar Sample and the Universal Mean/Median Composite Quasar SED}\label{sect:methods}
\unskip

\subsection{SDSS Quasars}\label{sect:sdss_quasars}

The SDSS DR14Q catalog contains 526,356 quasars \cite{Paris2018A&A...613A..51P}. Although $\simeq$$92\%$ of the DR14Q quasars are covered by the GALEX survey, the inhomogeneity of the GALEX survey,  with very diverse exposure times ranging from $\simeq$$30$ s to $\sim$$10^5$ s, makes the construction of {a meaningful} quasar SED very complex. The analysis of quasars covered by the GALEX tiles with long exposure times (e.g., $\geqslant$$10^3$ s; \cite{Cai2023NatAs}) should, to some extent, alleviate the problem.

In total, there are 45,195 GALEX {tiles}
\endnote{\url{https://galex.stsci.edu/Doc/gr67.txt}}. Dropping 311 tiles dedicated to the spectroscopic surveys, we consider 44,884 tiles for the imaging surveys. Nearly all (99.8\%) imaging tiles have positive exposure times in the GALEX near-UV (NUV) band, i.e., $t^{\rm tile}_{\rm NUV} > 1$ s, while 76.7\% have $t^{\rm tile}_{\rm FUV} > 1$ s in the GALEX far-UV (FUV) band. We require each GALEX tile to have positive exposure times in both GALEX bands {since we need the exposure times to determine flux upper limits for GALEX undetected quasars. Then, we} focus on 4688 imaging tiles with $t^{\rm tile}_{\rm FUV} \geqslant 10^3$ s, a trade-off between high GALEX detection and large sample size \cite{Cai2023NatAs}. 

Further requiring that the positions of the SDSS quasars are within the central $0.5^\circ$ radius of the field of view (FOV) of any one GALEX tile (that is, FOV offset $\leqslant 0.5^\circ$), we find that 107,504 SDSS quasars with positive flux densities and errors in all five SDSS bands are covered by 3453 GALEX tiles with $t^{\rm tile}_{\rm FUV} \geqslant 10^3$ s in the northern sky. As illustrated by the contours surrounded by sparse dots in the main panel of Figure \ref{fig:quasar_selection}, we consider the 101,745 SDSS quasars at $z \leqslant 3$ ($N^{\rm SDSS}_{\rm q}$; hereafter the parent quasar sample), beyond which both the GALEX detection is very low and the IGM absorption is rather significant (see below). 

Separating the parent quasar sample in redshift bins of 0.2 from $z = 0$ to $z = 3$, {the dotted histogram in the top panel of Figure \ref{fig:quasar_selection} shows the quasar number as a function of redshift, while the dotted histogram in the right panel of Figure \ref{fig:quasar_selection} shows the quasar number as a function of $\log L_{2200}$ for quasars in $1.8 < z \leqslant 2.0$ as an example. For each redshift bin, there is a specific value of $\log L_{2200}$ corresponding to the peak of the luminosity distribution of quasars (the right panel of Figure \ref{fig:quasar_selection}). These specific values of $\log L_{2200}$ for different redshifts constitute a red dashed stepwise curve in the main panel of Figure \ref{fig:quasar_selection}. Beneath the red dashed stepwise curve, the parent quasar sample is likely subject to significant sample incompleteness. Our unique bright quasar sample is selected well above the red dashed stepwise curve at $z > 0.5$, thus may not heavily suffer from the sample incompleteness.}

\subsection{GALEX Detections}\label{sect:galex_detections}

Searching the GALEX counterparts in the GALEX\_GR6Plus7 context\endnote{\url{https://galex.stsci.edu/casjobs/default.aspx}} for our parent quasars, we find 126,665 GALEX counterparts with positive exposure times in both GALEX bands within an SDSS/GALEX offset of $\leqslant$$2.6$ arcsec \cite{Trammell2007AJ....133.1780T}. Of the 101,745 parent quasars, $\simeq$$22.6\%$ have a unique counterpart, and $\simeq$$43.4\%$ have more than one counterpart, while the remaining $\simeq$$34.0\%$ do not have a counterpart. Following \citet{Cai2023NatAs}, a unique counterpart is selected for each quasar with multiple GALEX counterparts, while a $3\sigma$ GALEX NUV/FUV detection limit is assigned to each quasar either without a counterpart or with a unique counterpart but solely detected in the GALEX NUV (or FUV) band. 

\begin{figure}[H]
    \includegraphics[width=0.95\linewidth]{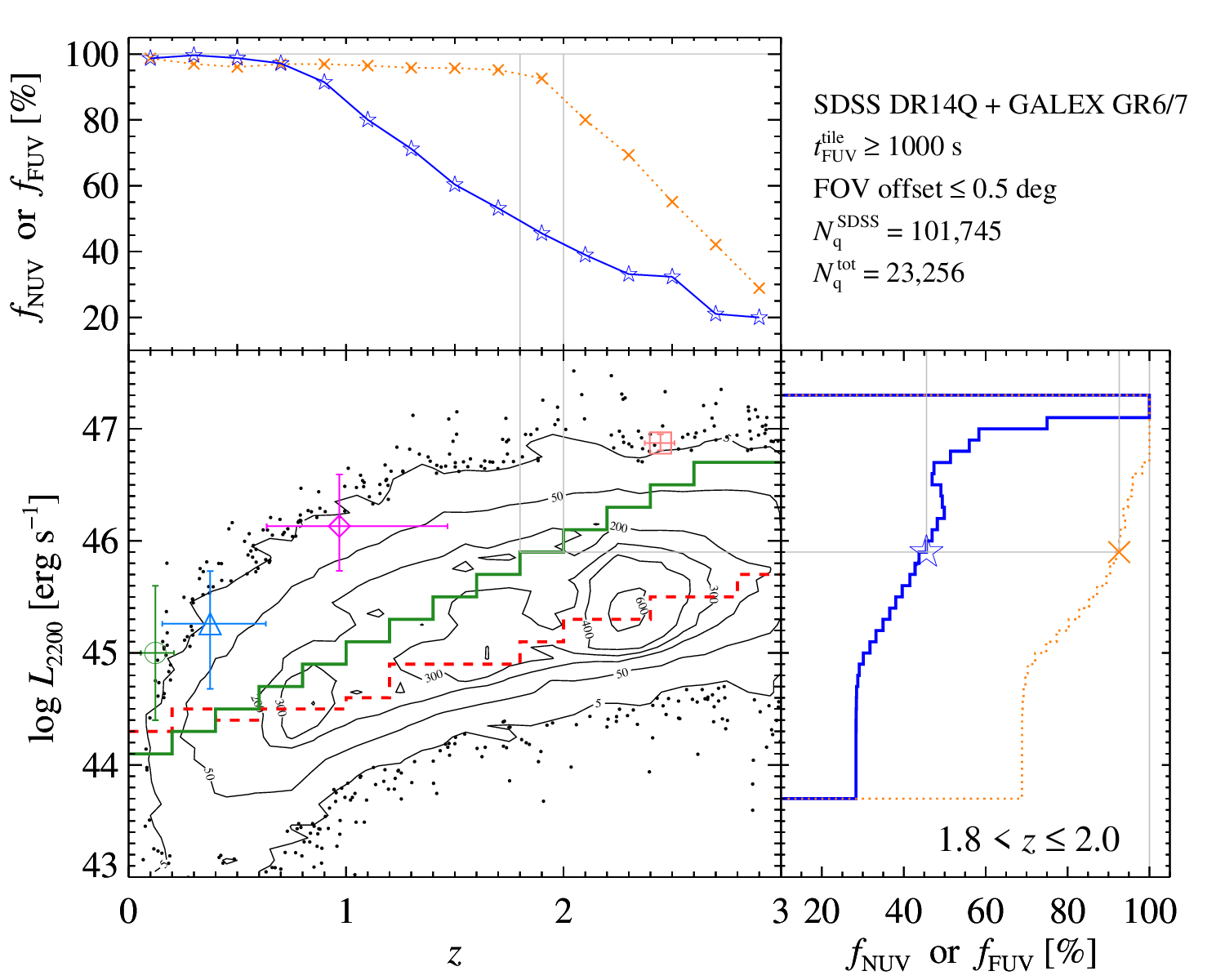}
    \caption{{The} 
 (\textbf{main}) panel is the same as that of Figure \ref{fig:quasar_selection}. The orange crosses linked by dotted lines (the blue stars linked by solid lines) in the (\textbf{top}) panel are the GALEX NUV (FUV) detection fractions, $f_{\rm NUV}$ ($f_{\rm FUV}$), for our unique bright quasars, while the orange dotted (blue solid) histogram in the (\textbf{right}) panel is the GALEX NUV (FUV) detection fraction as a function of luminosity for quasars brighter than a given minimal $2200 \angstrom$ luminosity and in $1.8 < z \leq 2.0$ as an example.
    }
    \label{fig:quasar_selection_uv_detection}
\end{figure}

In the right panel of Figure \ref{fig:quasar_selection_uv_detection}, the GALEX NUV- and FUV-detected fractions (i.e., $f_{\rm NUV}$ and $f_{\rm FUV}$) of our parent quasars in $1.8 < z \leqslant 2.0$ and brighter than a given minimal $2200 \angstrom$ luminosity, $\log L^{\rm min}_{2200}$, are shown as the orange dotted and blue solid histograms, respectively. It is clear that including fainter quasars gives rise to a larger sample size, but the GALEX detection fractions become lower. Therefore, for each redshift bin, it is critical to choose a specific luminosity limit such that both the quasar sample size and the GALEX detection fractions are not too small. \vspace{-6pt}

\begin{figure}[H]
    \includegraphics[width=\linewidth]{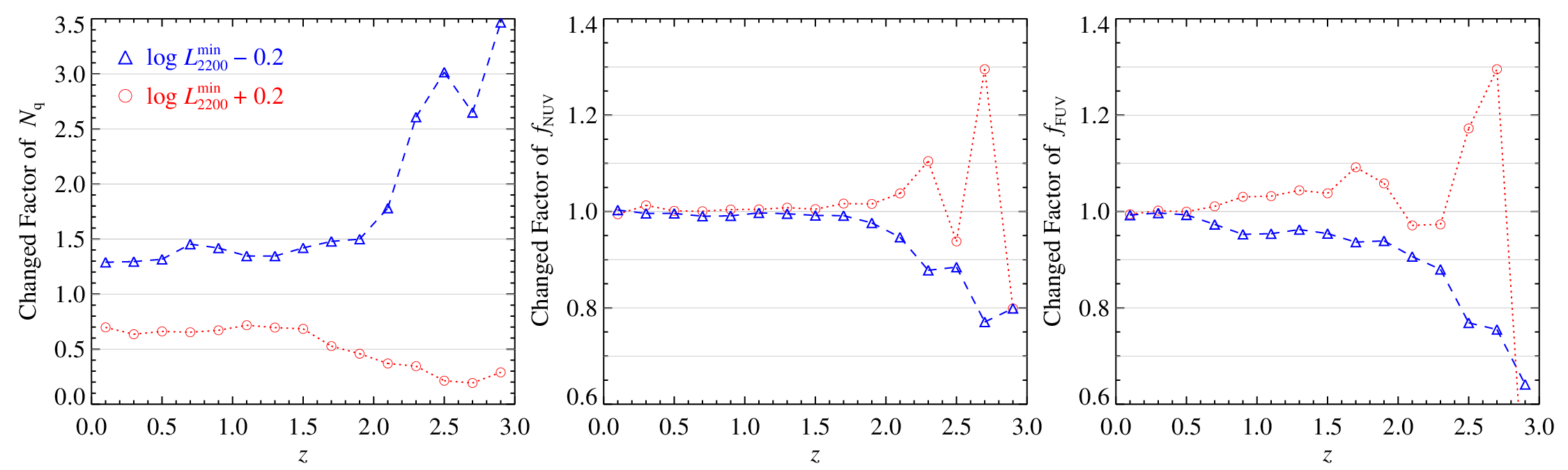}
     \caption{Changed factors of $N_{\rm q}$, $f_{\rm NUV}$, and $f_{\rm FUV}$ as a function of redshift for quasar samples selected with larger (circles linked by dotted lines) or smaller (triangles linked by dashed lines) $\log L_{2200}^{\rm min}$, i.e., by $\pm 0.2$ dex in all redshift bins, than our reference values of $\log L^{\rm min}_{2200}(z)$. Please note that the smaller $\log L_{2200}^{\rm min}$, the larger the sample size but the lower the GALEX detection.}
     \label{fig:nq_ffuv_fnuv_fit}
\end{figure}

Guided by the fact that quasars are typically brighter at higher redshift (Figure \ref{fig:quasar_selection_uv_detection}) and a luminosity limit of $\log L^{\rm min}_{2200} = 46$ has been used to construct a unique bright quasar sample at $z \sim 2$ by \citet{Cai2023NatAs}, we adopt redshift-dependent luminosity limits (the green solid stepwise curve in the main panel of Figures \ref{fig:quasar_selection} and \ref{fig:quasar_selection_uv_detection}), which increase from $\log L^{\rm min}_{2200} = 44.1$ for $0 < z \leqslant 0.2$ to $\log L^{\rm min}_{2200} = 46.7$ for $2.6 < z \leqslant 2.8$ in steps of 0.2 dex, except for the highest redshift bin where $\log L^{\rm min}_{2200} = 46.7$ is adopted to avoid including too few quasars. The top panel of Figure \ref{fig:quasar_selection} shows the number of quasars brighter than $\log L^{\rm min}_{2200}$ as a function of redshift (green solid histogram), while the top panel of Figure \ref{fig:quasar_selection_uv_detection} illustrates the corresponding $f_{\rm NUV}$ (blue open stars) and $f_{\rm FUV}$ (orange crosses) as a function of redshift. In the following, our mean/median composite quasar SED would be constructed using 23,256 quasars brighter than $\log L^{\rm min}_{2200}$ (hereafter, the unique bright quasar sample).
{We note that utilizing other quasar samples selected with larger or smaller $\log L^{\rm min}_{2200}$, e.g., by $\pm 0.2$ dex in all redshift bins, than our reference values of $\log L^{\rm min}_{2200}(z)$, confirms our conclusion on the universality of the quasar SED. However, the EUV portion of the resultant SED is more uncertain owing to either smaller sample size for larger $\log L^{\rm min}_{2200}$ or lower GALEX detection fraction for smaller $\log L^{\rm min}_{2200}$ (Figure \ref{fig:nq_ffuv_fnuv_fit} and Section \ref{sect:universal_seds}).}

\unskip

\subsection{Bias-Free Mean/Median Quasar SEDs in Different Redshift Bins}

After considering the GALEX non-detections, we construct the bias-free average (i.e., both mean and median) SEDs for quasars in different redshift bins and brighter than the corresponding $\log L^{\rm min}_{2200}$. A brief introduction to the method is given here, while readers are referred to \citet{Cai2023NatAs} for all details. 

For each quasar in a redshift bin, its rest-frame SED consists of de-redshifted SDSS {($ugriz$)} 
 and GALEX (FUV and NUV) photometry, where the GALEX non-detections are replaced by $3\sigma$ upper detection limits. Each quasar SED is normalized at the rest-frame $2200 \angstrom$ before taking the mean/median value of all normalized SEDs. The mean/median quasar SED is easily reachable in the de-redshifted SDSS bands, and the $1\sigma$ statistical errors are estimated as the standard deviation of all normalized SEDs divided by the square root of the quasar number in each redshift bin \cite{Trammell2007AJ....133.1780T}. Instead, since there are non-detections in the de-redshifted GALEX bands, an exponentially modified Gaussian function \cite{VandenBerk2020MNRAS.493.2745V} is fit to the distribution of $\log(L_w/L_{2200})$, where $L_w$ is the rest-frame $w \angstrom$ monochromatic luminosity. Here, the subscript $w$ indicates the rest-frame wavelength corresponding to the effective wavelength of the GALEX NUV/FUV band divided by $1+z$, where $z$ is the median redshift of quasars in each redshift bin. Then, the mean/median quasar SED at $w \angstrom$ is inferred from the best-fit exponentially modified Gaussian function, while the $1\sigma$ uncertainty is estimated by bootstrapping 1000 times the distribution of $\log(L_w/L_{2200})$, including both GALEX detections and non-detections.

\begin{figure}[H]
\hspace{-1em}    \includegraphics[width=0.9\linewidth]{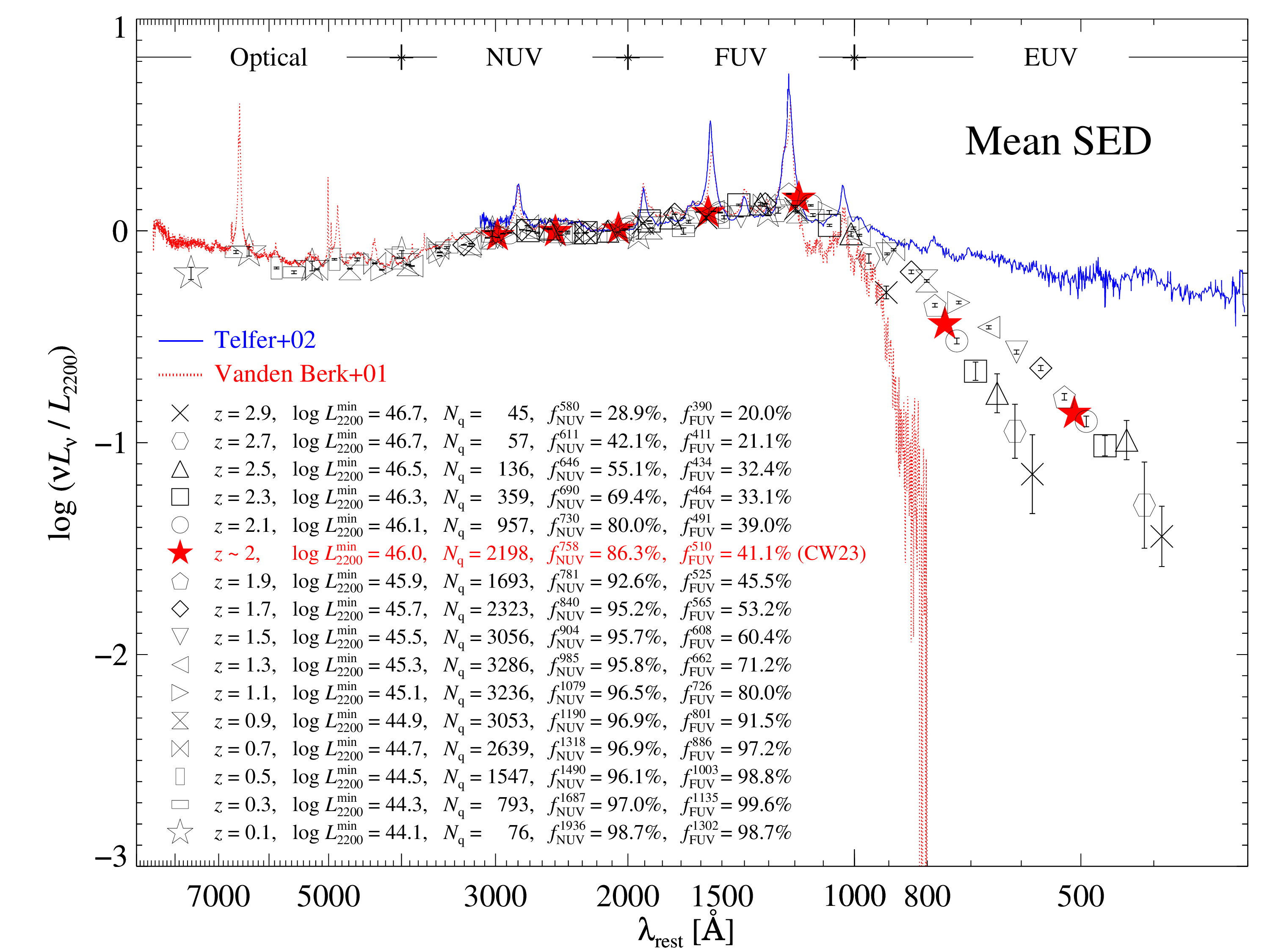}

  \hspace{-1em}  \includegraphics[width=0.9\linewidth]{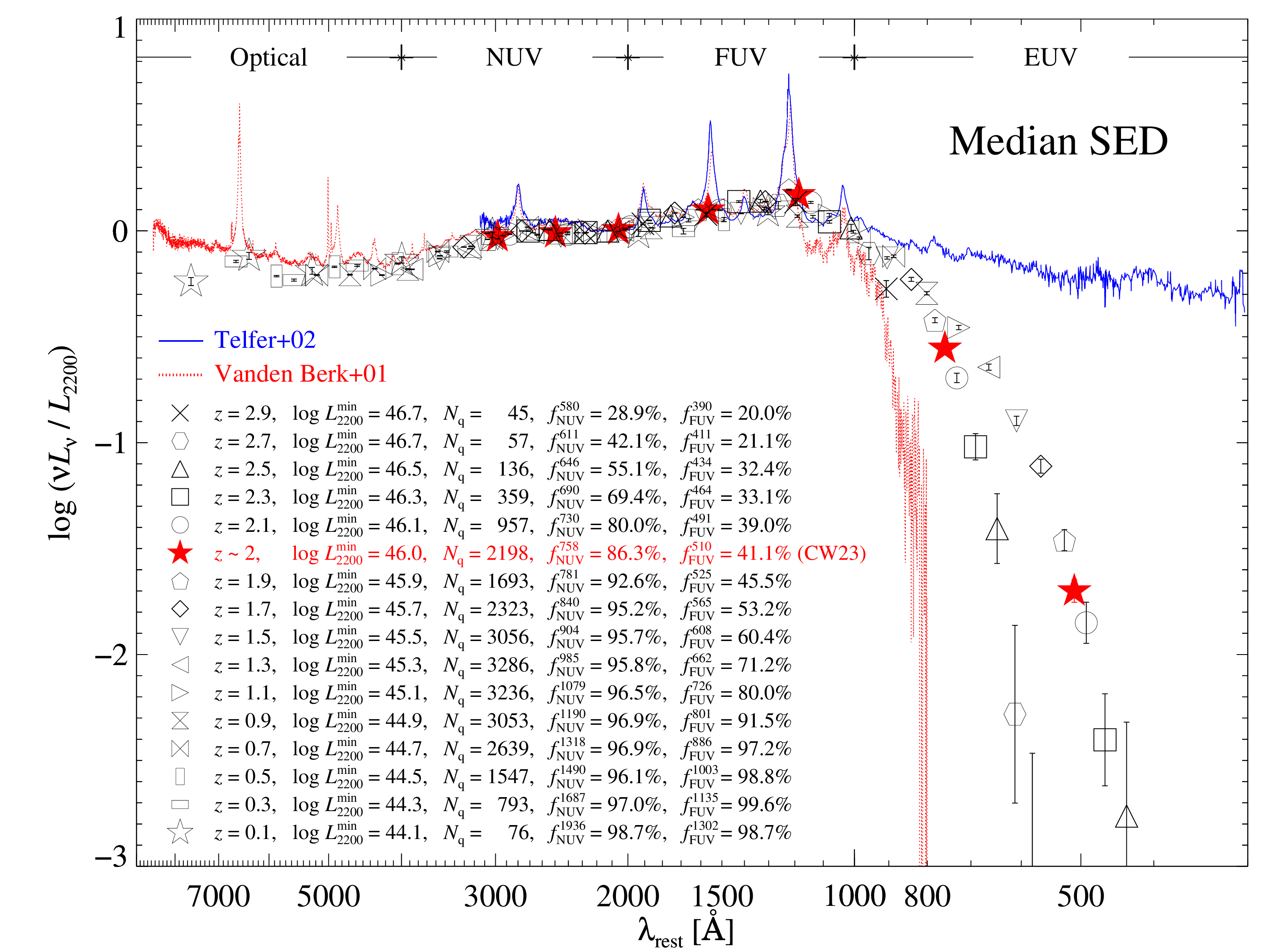}
     \caption{The (\textbf{top}) {panel:} 
 the bias-free mean SED for $N_{\rm q}$ quasars brighter than $\log L^{\rm min}_{2200}$ in each redshift bin. Each legend contains the GALEX NUV- and FUV-detected fractions, i.e., $f^w_{\rm NUV}$ and $f^w_{\rm FUV}$, where the superscript, $w$, is the de-redshifted wavelength corresponding to the effective wavelength of the GALEX NUV/FUV band. 
     We also show the mean quasar SED constructed by (\cite{Cai2023NatAs}, CW23) as well as the mean composite quasar spectra from (\cite{VandenBerk2001AJ....122..549V}, without correction for the IGM absorption; red dotted curve) and (\cite{Telfer2002ApJ}, with correction for the IGM absorption; blue solid curve) for comparison.
     The (\textbf{bottom}) panel: same as the (\textbf{top}) one, but for the bias-free median quasar SEDs.
     }
     \label{fig:quasar_sed_bias_free}
\end{figure}

Figure \ref{fig:quasar_sed_bias_free} shows the resultant mean (top panel) and median (bottom panel) SEDs for quasars in redshift bins with central redshifts increasing from $z = 0.1$ to $z = 2.9$ in steps of 0.2. Both the mean and median SEDs at the optical-to-FUV wavelengths beyond $\simeq$$1000 \angstrom$ are in good agreement with previous mean composite spectra, such as \citet{VandenBerk2001AJ....122..549V} and \citet{Telfer2002ApJ}. The consistency between the mean and median optical-to-FUV SEDs indicates that distributions of the optical-to-FUV luminosities are symmetric. 
Instead, both the mean and median quasar EUV SEDs seemingly become redder with increasing redshift. Since the IGM absorption is also more significant with increasing redshift, these bias-free SEDs of quasars at different redshifts, without corrections for the IGM absorption, cannot be compared to each other and to the EUV portion of the mean composite spectrum of \citet{Telfer2002ApJ}, to which corrections for the IGM absorption has been applied.

\unskip

\subsection{Monte Carlo Simulation for the IGM Absorption and Correction}

\begin{figure}[H]
    \includegraphics[width=0.5\linewidth]{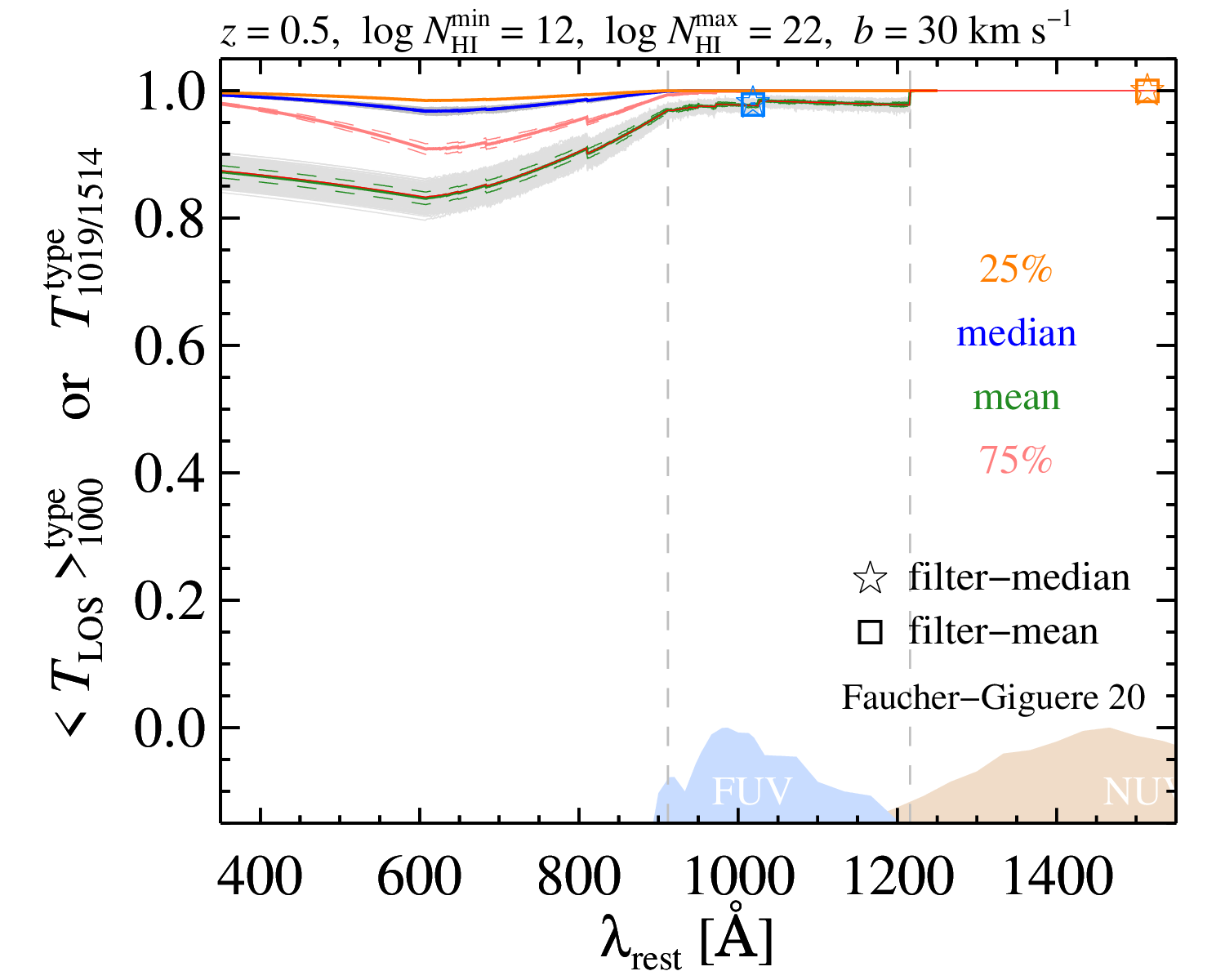}
     \includegraphics[width=0.5\linewidth]{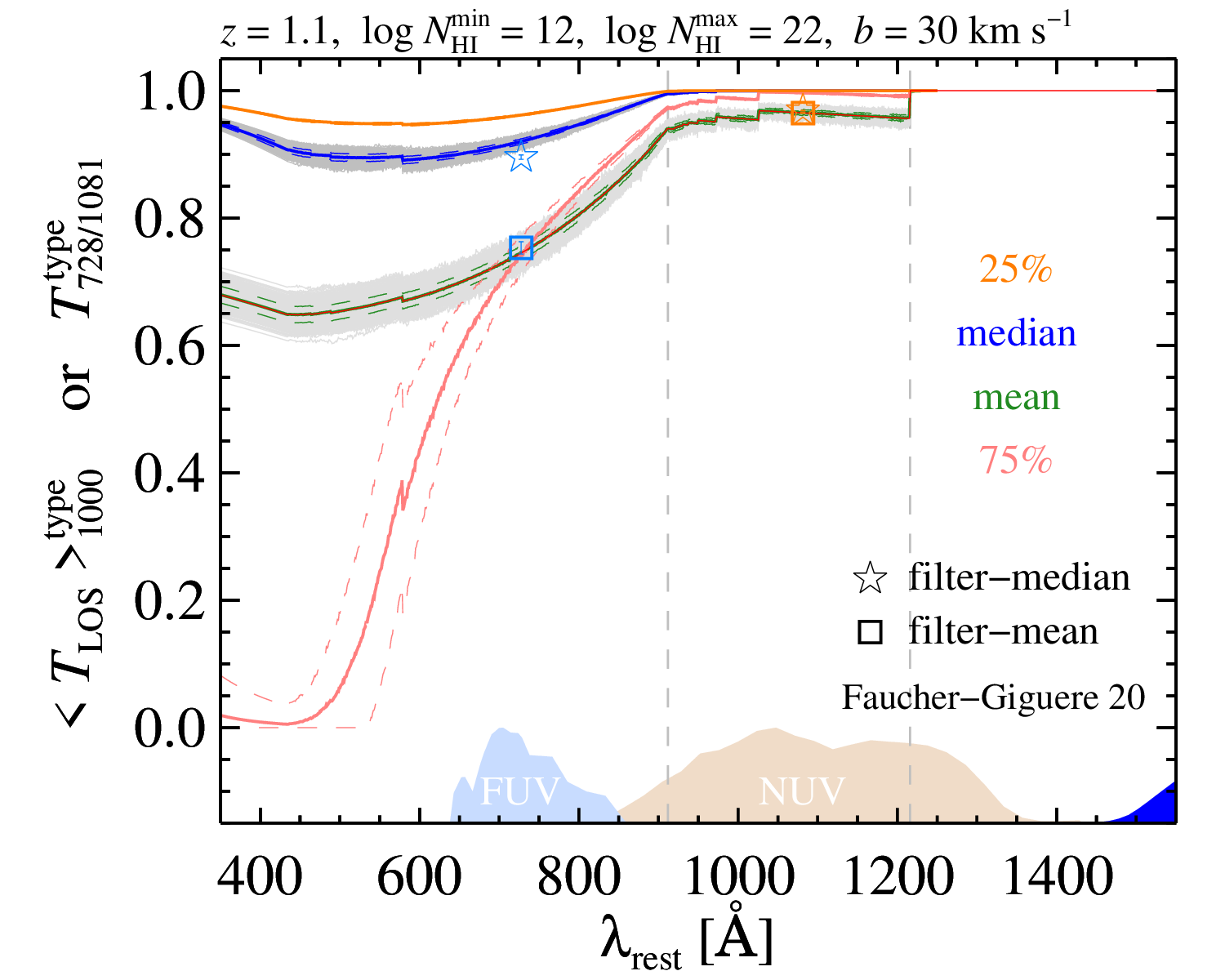}
      \includegraphics[width=0.5\linewidth]{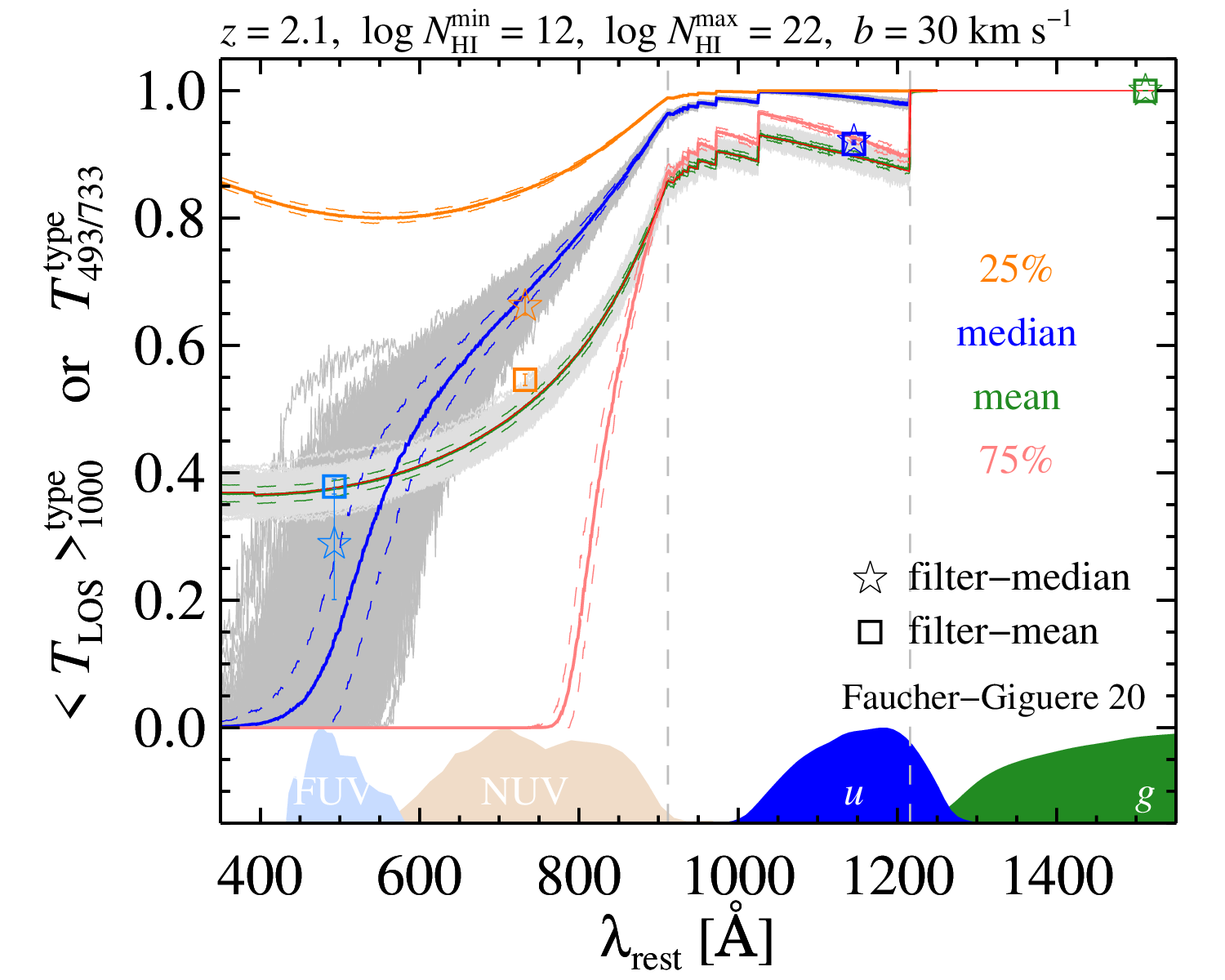}
       \includegraphics[width=0.5\linewidth]{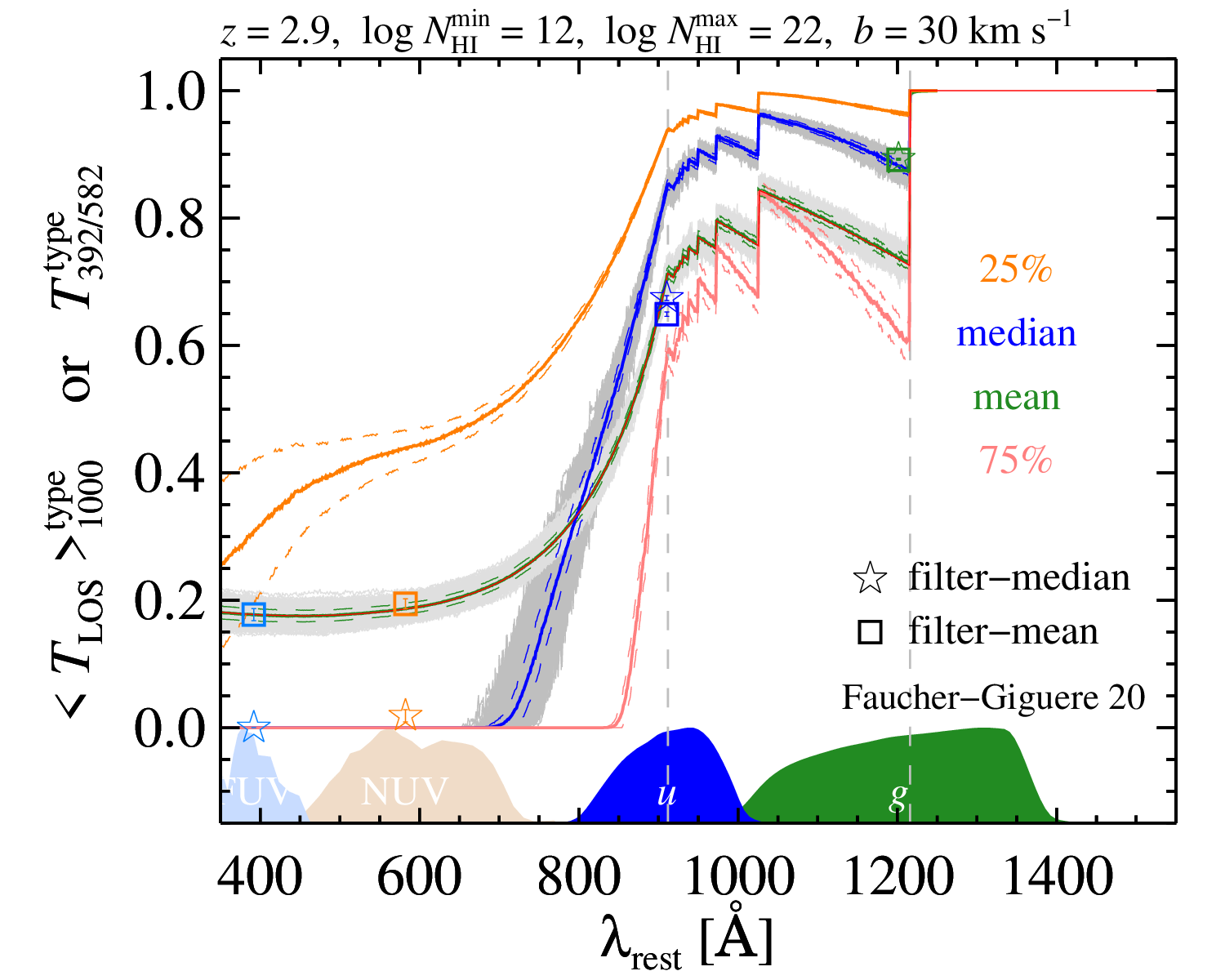}
     \caption{{The} 
 IGM transmission curves as a function of the rest-frame wavelength for $z = 0.5$ (\textbf{top-left} panel), $z = 1.1$ (\textbf{top-right} panel), $z = 2.1$ (\textbf{bottom-left} panel), and $z = 2.9$ (\textbf{bottom-right} panel). Averaging transmissions in 1000 LOS (equivalent to 1000 quasars) randomly selected from $10^5$ simulated LOS gives rise to the mean transmission and three upper transmission quantiles (i.e., 25\%, median, and 75\%), $\langle T_{\rm LOS} \rangle_{1000}^{\rm type}$. For each ``type'' of transmission, the solid and dashed curves are the mean and $1\sigma$ dispersion of 1000 different realizations of $\langle T_{\rm LOS} \rangle_{1000}^{\rm type}$ (cf. the gray and light-gray curves for 1000 realizations of $\langle T_{\rm LOS} \rangle_{1000}^{\rm median}$ and $\langle T_{\rm LOS} \rangle_{1000}^{\rm mean}$, respectively). At the bottom of each panel, there are de-redshifted FUV-, NUV-, $u$-, and $g$-band transmission curves from left to right, respectively. In each panel and each band, the square and star, superimposed by $1\sigma$ uncertainties, indicate the filter-weighted broadband mean and median IGM transmissions, respectively.}
     \label{fig:igm_correction}
\end{figure}

Details on the Monte Carlo simulation for the IGM transmission curves are documented in \citet{Cai2023NatAs}. {Adopting the \citet{Faucher-Giguere2020MNRAS.493.1614F} distribution of absorbers with the \HI~column density of $12 \leqslant \log N_{\mbox{\scriptsize \HI}} \leqslant 22$ and the Doppler broadening parameter of $b = 30 {\rm km s^{-1}}$,} Figure \ref{fig:igm_correction} illustrates the average IGM transmission curves for 1000 quasars at four different redshifts. Globally, the IGM absorption is more significant with increasing redshift. At all redshifts, the mean IGM transmission curve is very different from the median IGM transmission curve. At low redshifts, e.g., $z = 0.5$, although less than $25\%$ quasars are heavily absorbed owing to the rare encounter with Lyman limit systems along the LOS to quasars, the mean IGM transmission curve is somewhat lower than the median IGM transmission curve. Instead, at high redshifts, e.g., $z = 2.9$, there are still $\sim$$25\%$ quasars that {are not} heavily absorbed such that the mean IGM transmission at $<$$700 \angstrom$ remains about $\sim$$0.2$ when the median IGM transmission is almost zero.

\newpage
In practice, the filter-weighted broadband mean and median IGM transmissions are used to correct the IGM-absorbed EUV portion of the bias-free mean and median SEDs, respectively. If there are $N_{\rm q}$ quasars in a redshift bin with a central redshift of $z$, the mean/median broadband IGM transmission is estimated assuming all $N_{\rm q}$ quasars are at $z$, regardless of the slight differences in redshift among quasars. 
{We obtain a value for the mean/median IGM transmission by averaging $N_{\rm q}$ LOS randomly selected from $10^5$ LOS, repeat the random selection 1000 times, and take the standard deviation of the 1000 values as the $1\sigma$ uncertainty for the mean/median broadband IGM transmission given $N_{\rm q}$ quasars. Figure \ref{fig:igm_correction_broadband} illustrates the mean/median broadband IGM transmissions as well as the $1\sigma$ uncertainties for 1000 quasars at each redshift as an example. Finally, the applied $1\sigma$ uncertainty of the mean/median broadband IGM correction is estimated according to the real quasar number in each redshift bin, and} is propagated to the EUV portion of the intrinsic mean/median quasar SED.
\vspace{-6pt}

\begin{figure}[H]
 \hspace{-0.5em}   \includegraphics[width=1.0\linewidth]{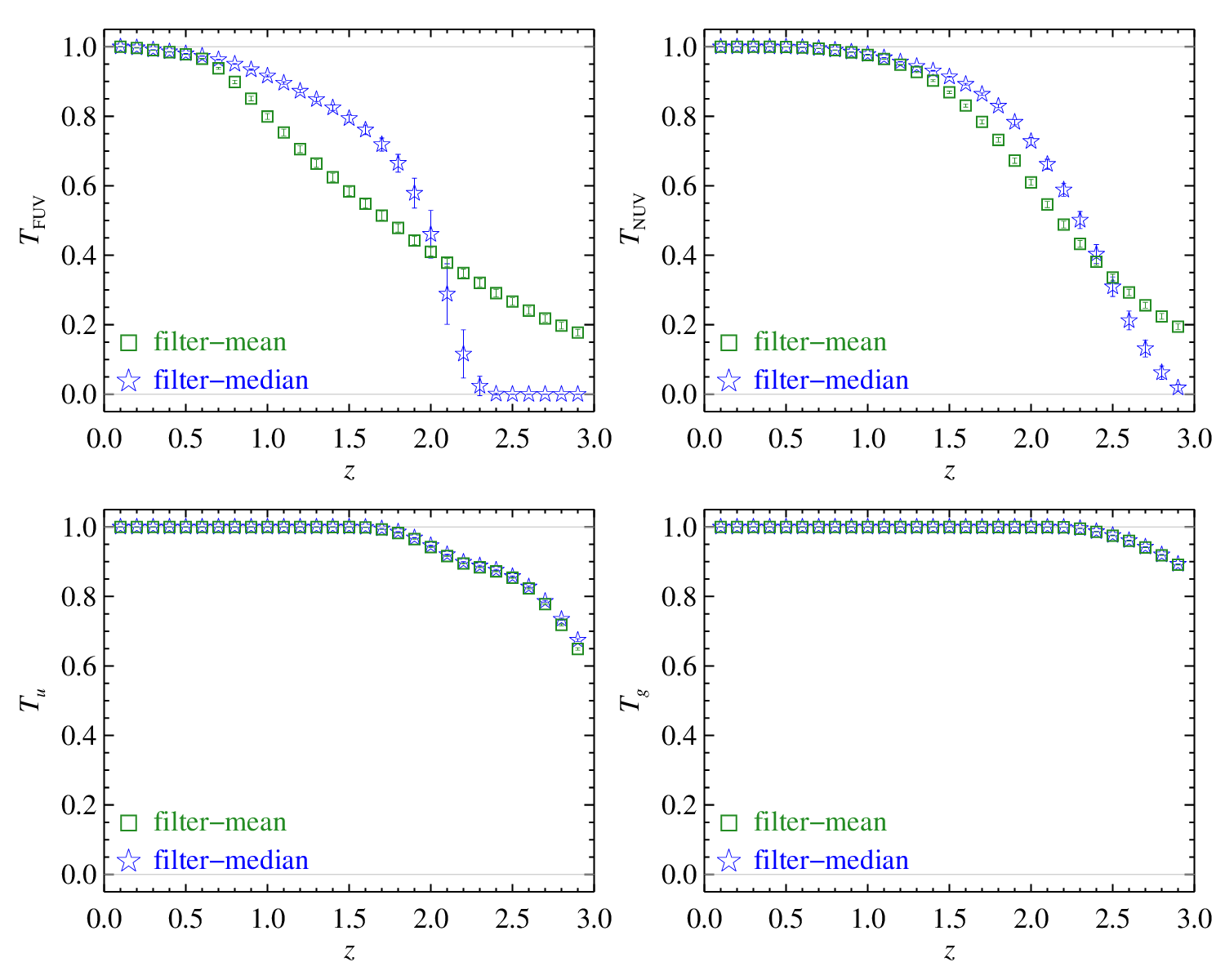}
     \caption{By averaging 1000 LOS, squares and stars show the filter-weighted broadband mean and median IGM transmission as a function of redshift for four bands, i.e., $T_{\rm FUV}$ (\textbf{top-left}), $T_{\rm NUV}$ (\textbf{top-right}), $T_{u}$ (\textbf{bottom-left}), and $T_{g}$ (\textbf{bottom-right}). The superimposed $1\sigma$ uncertainties are associated with 1000 LOS. Please note that averaging more (less) LOS gives smaller (larger) $1\sigma$ uncertainties.}
     \label{fig:igm_correction_broadband}
\end{figure}
\unskip
\subsection{An Intrinsic Mean/Median Composite SED for Quasars Since Cosmic Noon}

After applying corrections for the IGM absorption, Figure \ref{fig:quasar_sed_composite} illustrates the intrinsic mean (top panel) and median (bottom panel) SEDs for quasars at different redshifts. In contrast to the bias-free quasar SEDs in the EUV (Figure \ref{fig:quasar_sed_bias_free}), the EUV portions of these intrinsic quasar SEDs are strikingly consistent with each other. The EUV portions of the intrinsic quasar SEDs are contributed by quasars at $z \gtrsim 0.5$ up to $z \sim 2.9$ and with typical $2200 \angstrom$ luminosities increasing from $\log L^{\rm med}_{2200} = 44.8$ to $\log L^{\rm med}_{2200} = 46.8$. Consequently, these intrinsic mean/median quasar SEDs strongly suggest an intrinsic mean/median composite SED for quasars since cosmic noon.
\newpage

The intrinsic median quasar SED at the rest-frame $\lambda_{\rm rest} < 500 \angstrom$ is very uncertain because of both the low GALEX detection fractions, smaller than 35\% (Figures \ref{fig:quasar_selection_uv_detection} and \ref{fig:quasar_sed_bias_free}), and the significantly uncertain corrections for the broadband median IGM transmissions (top panels of Figure \ref{fig:igm_correction_broadband}). However, the intrinsic mean quasar SED at $\lambda_{\rm rest} < 500 \angstrom$ is plausible since (1) the mean value is not sensitive to those small values corresponding to the upper limits assigned for the GALEX non-detections and (2) the rather reliable corrections for the broadband mean IGM transmissions (top panels of Figure \ref{fig:igm_correction_broadband}).

The intrinsic mean/median composite quasar SED is fit by a smoothly broken power law in the form of
\begin{equation}
    \nu L_\nu \propto \left[ \left(\frac{\lambda}{\lambda_{\rm b}}\right)^{{\cal S} (1 + \alpha_{\rm OPT-FUV}) {\cal C}} + \left(\frac{\lambda}{\lambda_{\rm b}}\right)^{{\cal S} (1+\alpha_{\rm EUV}) {\cal C}} \right]^{{\cal S}/{\cal C}},
\end{equation}
where $\lambda_{\rm b}$ is the break wavelength, $\alpha_{\rm OPT-FUV}$ is the optical-to-FUV spectral index at $\lambda \gg \lambda_{\rm b}$, $\alpha_{\rm EUV}$ is the EUV spectral index at $\lambda \ll \lambda_{\rm b}$, $\cal C$ is the curvature parameter, and $\cal S$ is the sign of ($\alpha_{\rm EUV} - \alpha_{\rm OPT-FUV}$). 
{We consider} the following line-free windows for the continuum fitting: $480$--$1000$, $1060$--$1140$, $1350$--$1450$, $1700$--$1800$, $2000$--$2300$, $3800$--$4200$, and $5100$--$6000 \angstrom$ (gray thick bars in Figure \ref{fig:quasar_sed_composite}).
A minimum $\chi^2$ approach performed using the routine {\texttt{MPFIT}}\endnote{\url{https://cow.physics.wisc.edu/ craigm/idl/idl.html}} gives the best-fit parameters and the associated $1\sigma$ uncertainties.
For the intrinsic mean composite quasar SED, we find that $\alpha_{\rm OPT-FUV} = -0.535 \pm 0.003$, $\alpha_{\rm EUV} = -2.73 \pm 0.03$, $\lambda_{\rm b} = 1144.0 \pm 6.2 \angstrom$, and ${\cal C} = 9.34 \pm 1.73$.
For the intrinsic median composite quasar SED, we find that $\alpha_{\rm OPT-FUV} = -0.468 \pm 0.003$, $\alpha_{\rm EUV} = -6.78 \pm 0.21$, $\lambda_{\rm b} = 906.0 \pm 11.4 \angstrom$, and ${\cal C} = 1.08 \pm 0.07$.

\begin{figure}[H]
 \hspace{-1.5em}   \includegraphics[width=1\linewidth]{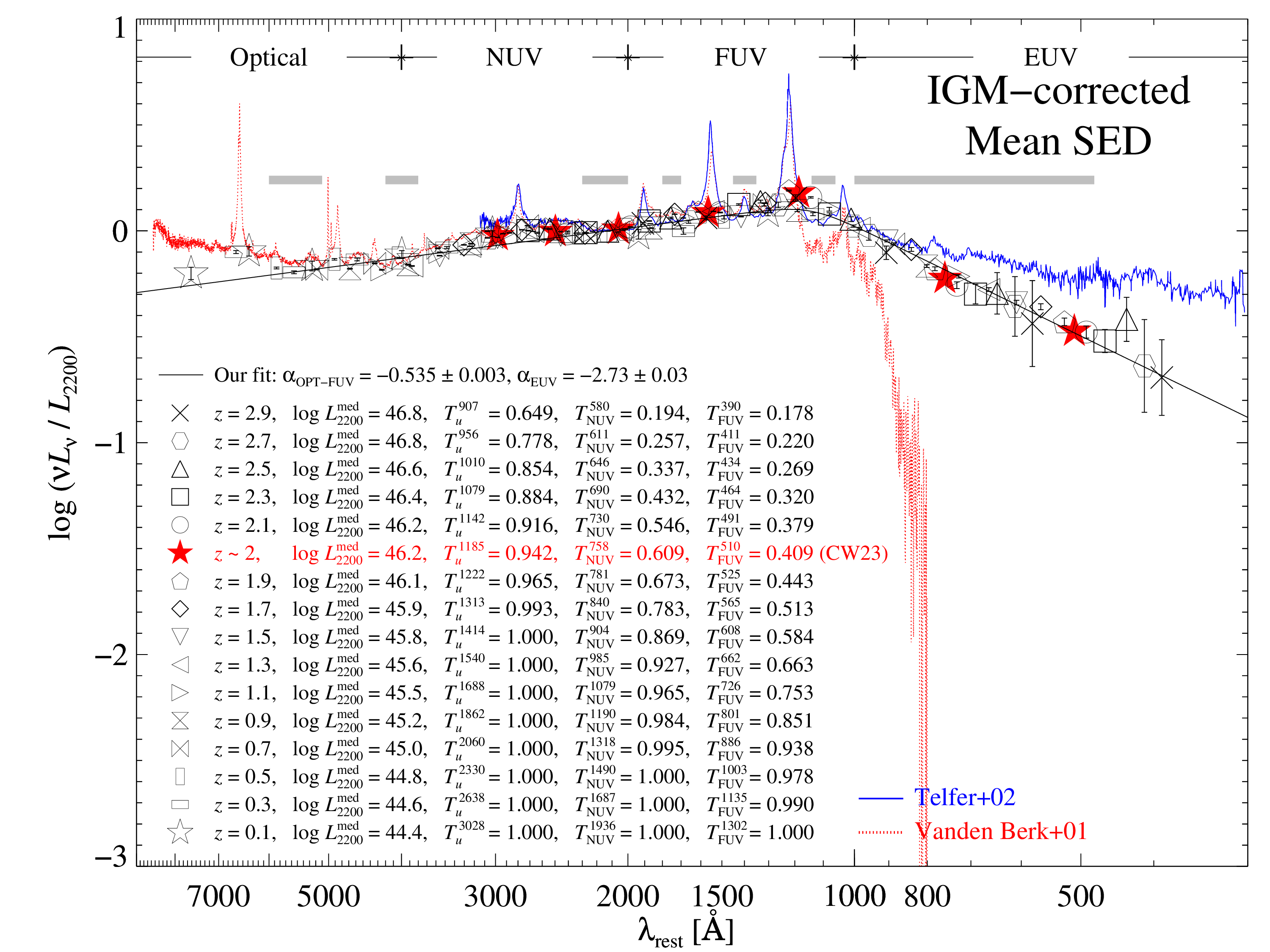}
\caption{{\em Cont.}}
     \label{fig:quasar_sed_composite}
\end{figure}

\begin{figure}[H]\ContinuedFloat
 \hspace{-1.5em}   \includegraphics[width=1\linewidth]{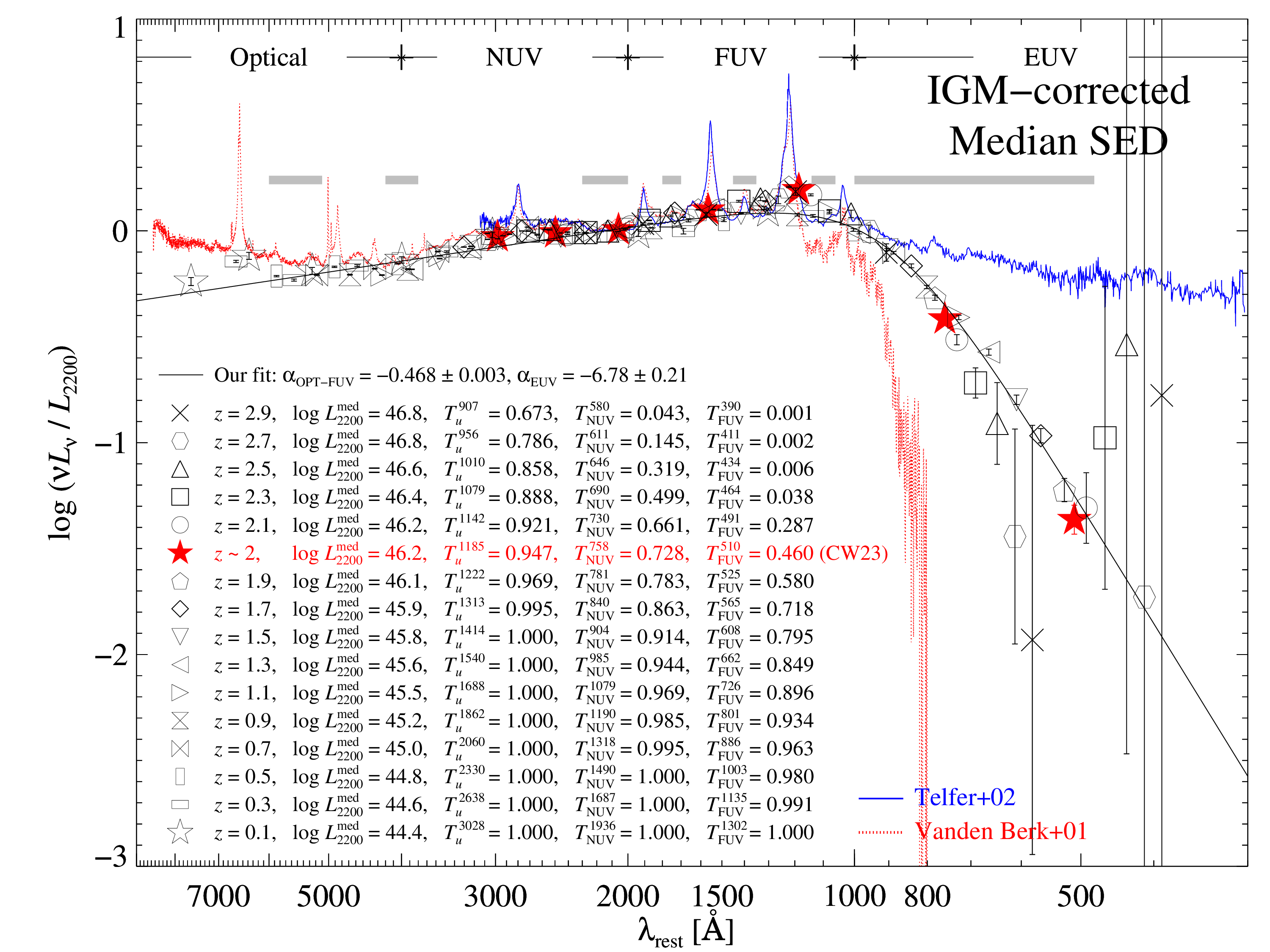}
     \caption{{Same} 
 as Figure \ref{fig:quasar_sed_bias_free}, but for the intrinsic mean (\textbf{top} panel) and median (\textbf{bottom} panel) SEDs, whose EUV portion has been corrected for the IGM absorption. For each redshift bin with a central redshift $z$, the legend includes the median $2200 \angstrom$ luminosity of quasars, $\log L^{\rm med}_{2200}$, and the filter-weighted broadband IGM transmissions, $T^w_{b}$, where $b$ indicates the band name. After correcting for the IGM absorption, the intrinsic mean/median SEDs of quasars at different redshifts become strikingly consistent in the EUV and together form a mean/median composite SED. A smoothly broken power law with optical-to-FUV ($\alpha_{\rm OPT-FUV}$) and EUV ($\alpha_{\rm EUV}$) spectral indices for $L_\nu \propto \nu^{\alpha}$ is fit to the mean/median composite SED in several line-free windows (thick gray bars).}
     \label{fig:quasar_sed_composite}
\end{figure}
\unskip

\section{Discussion and Implications}\label{sect:discussion}
\unskip

\subsection{Universality of the Mean/Median Composite Quasar SED}\label{sect:universal_seds}

After controlling the EUV detection incompleteness bias for a quasar sample at $z \sim 2$, \citet{Cai2023NatAs} suggested that there is a universal luminosity-independent average SED for quasars at least for $\log L_{2200} > 45$. The claim was made in terms of a small number of quasars at $z \sim 2$, and more critically, only $\simeq 3\%$ (up to $20\%$) of them were used in order to compare the same fraction of the brightest quasars among different luminosity bins. 
Here, the consistency among the intrinsic average SEDs of quasars at different redshifts strongly strengthens our previous claim on the luminosity independence of the average quasar SED. 

\begin{figure}[H]
    \includegraphics[width=1.0\linewidth]{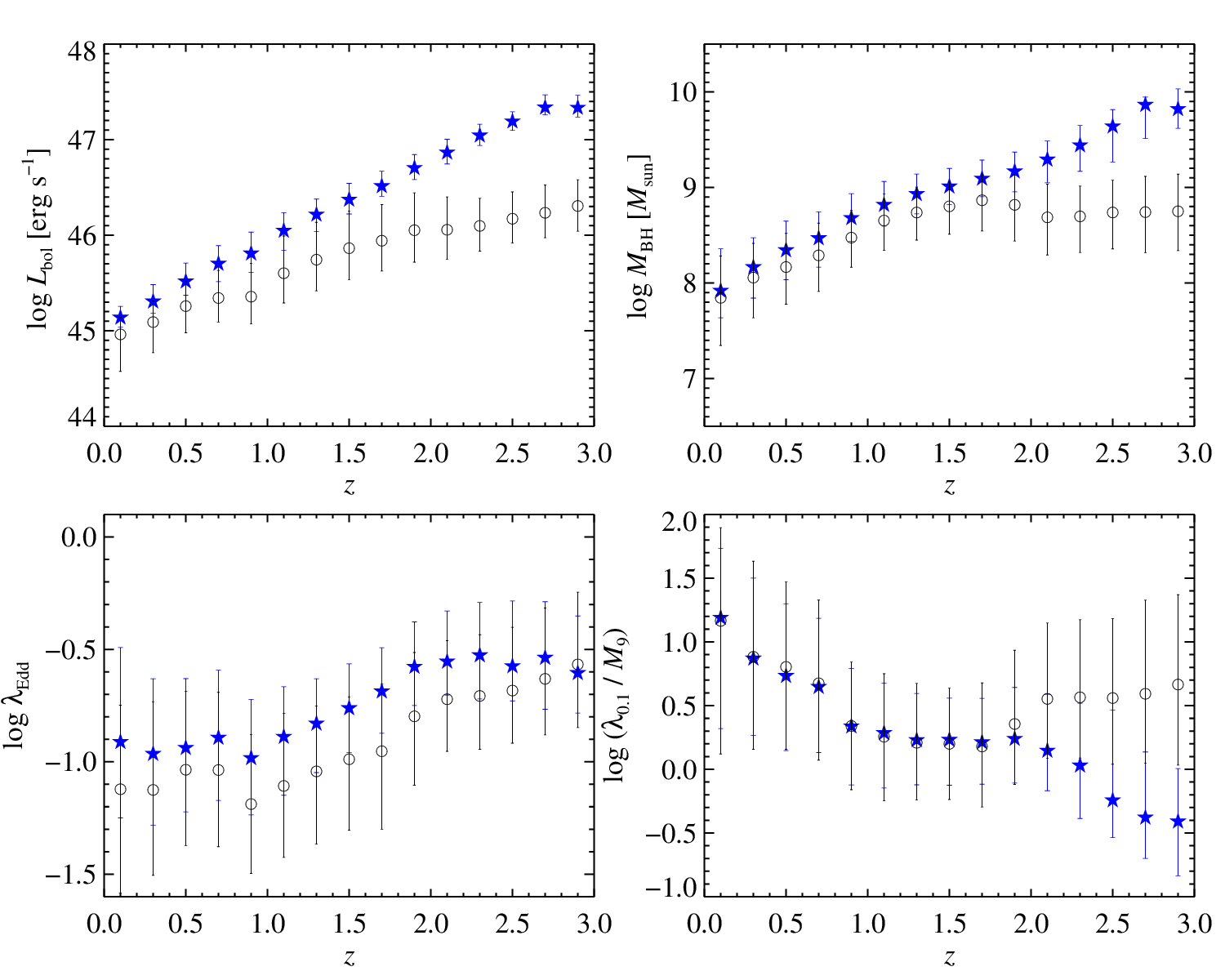}
     \caption{Physical properties, i.e., bolometric luminosity ($L_{\rm bol}$; \textbf{top-left} panel), BH mass ($M_{\rm BH}$; \textbf{top-right} panel), Eddington ratio ($\lambda_{\rm Edd}$; \textbf{bottom-left} panel), and $\lambda_{\rm Edd} / M_{\rm BH}$ (\textbf{bottom-right} panel; note $\lambda_{0.1} = \lambda_{\rm Edd} / 0.1$ and $M_9 = M_{\rm BH} / 10^9 M_\odot$), as a function of redshift for our parent quasar sample (black open circles; Section \ref{sect:sdss_quasars}) and our unique bright quasar sample (blue filled stars; Section \ref{sect:galex_detections}). 
     At each redshift, symbols are median values, while vertical bars are the 25–75th percentile ranges. 
     Nearly all quasars have measurements on $L_{\rm bol}$, $M_{\rm BH}$, and $\lambda_{\rm Edd}$, provided by \citet{Rakshit2020ApJS..249...17R}, but the large uncertainties on, and even systematic offsets of, these measurements must be taken seriously (see discussion in \citet{Sun2023MNRAS.521.2954S} for example).}
     \label{fig:physical_properties_z}
\end{figure}

For $z > 0.5$ quasars, capable of providing EUV measurements with the help of GALEX, the median $2200 \angstrom$ luminosity increases from $\log L^{\rm med}_{2200} = 44.8$ for $z \sim 0.5$ quasars to $\log L^{\rm med}_{2200} = 46.8$ for $z \sim 2.9$ quasars. Regardless of the uncertainties in the conversion of the monochromatic optical/UV luminosity into the bolometric luminosity by adopting a redshift-dependent bolometric correction factor \cite{Rakshit2020ApJS..249...17R}, the median bolometric luminosity increases $\log L_{\rm bol} \simeq 45.5$ for $z \sim 0.5$ quasars to $\log L_{\rm bol} \simeq 47.3$ for $z \sim 2.9$ quasars (top-left panel of Figure \ref{fig:physical_properties_z}). Therefore, over two orders of magnitude in luminosity and since cosmic noon, our universal mean/median composite quasar SED demonstrates not only the luminosity independence but also the redshift independence of {the average optical-to-EUV radiation properties of quasars}. 

Moreover, since both $M_{\rm BH}$ and $\lambda_{\rm Edd}$ of our unique bright quasar sample used to construct the mean/median composite quasar SED are dependent on redshift, the universality of our composite quasar SED further implies an independence on both BH mass and Eddington ratio, at least for quasars with $M_{\rm BH} \gtrsim 10^8 M_\odot$ and $\lambda_{\rm Edd} \gtrsim 0.1$ (Figure \ref{fig:physical_properties_z}). However, for low-luminosity AGN with $\lambda_{\rm Edd} \sim 0.01$, the universality may break down, given a potential transition of the accretion modes \cite{Taam2012ApJ...759...65T,Hagen2024MNRAS.tmp.2224H,Kang2024arXiv241006730K}.

\begin{figure}[H]
\begin{adjustwidth}{-\extralength}{0cm}
\centering 

    \includegraphics[width=0.42\linewidth]{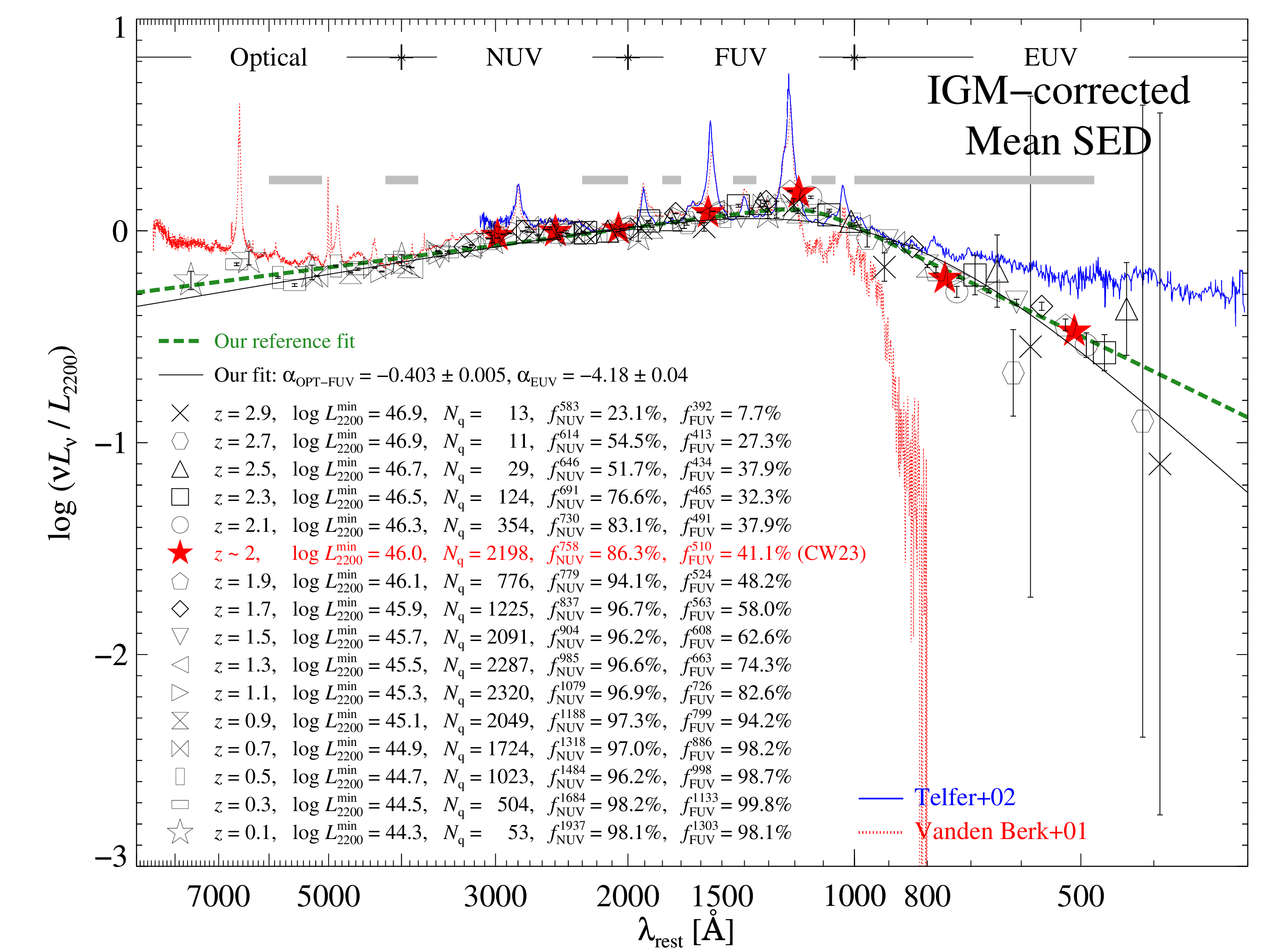}
    \includegraphics[width=0.42\linewidth]{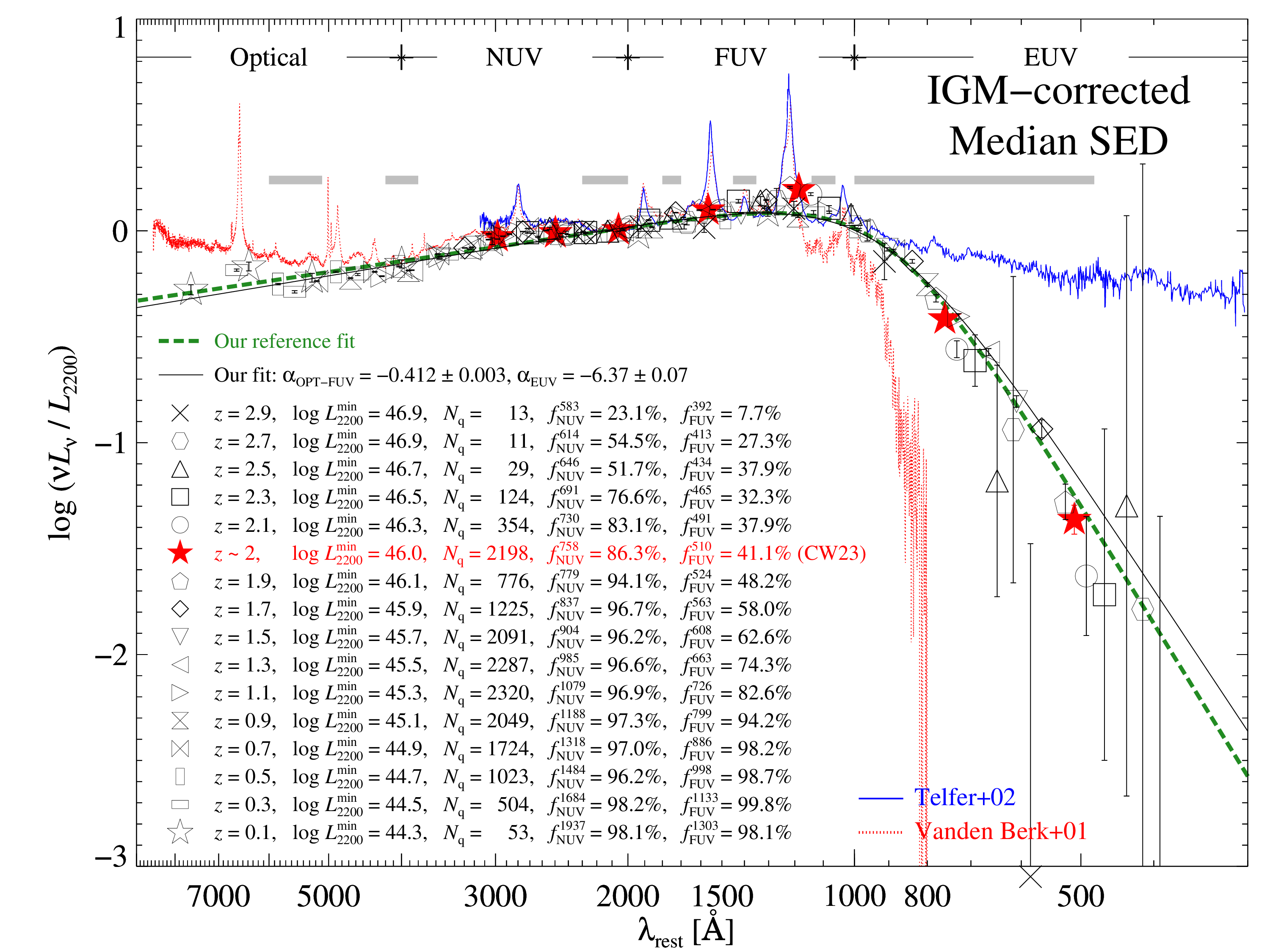}

    \includegraphics[width=0.42\linewidth]{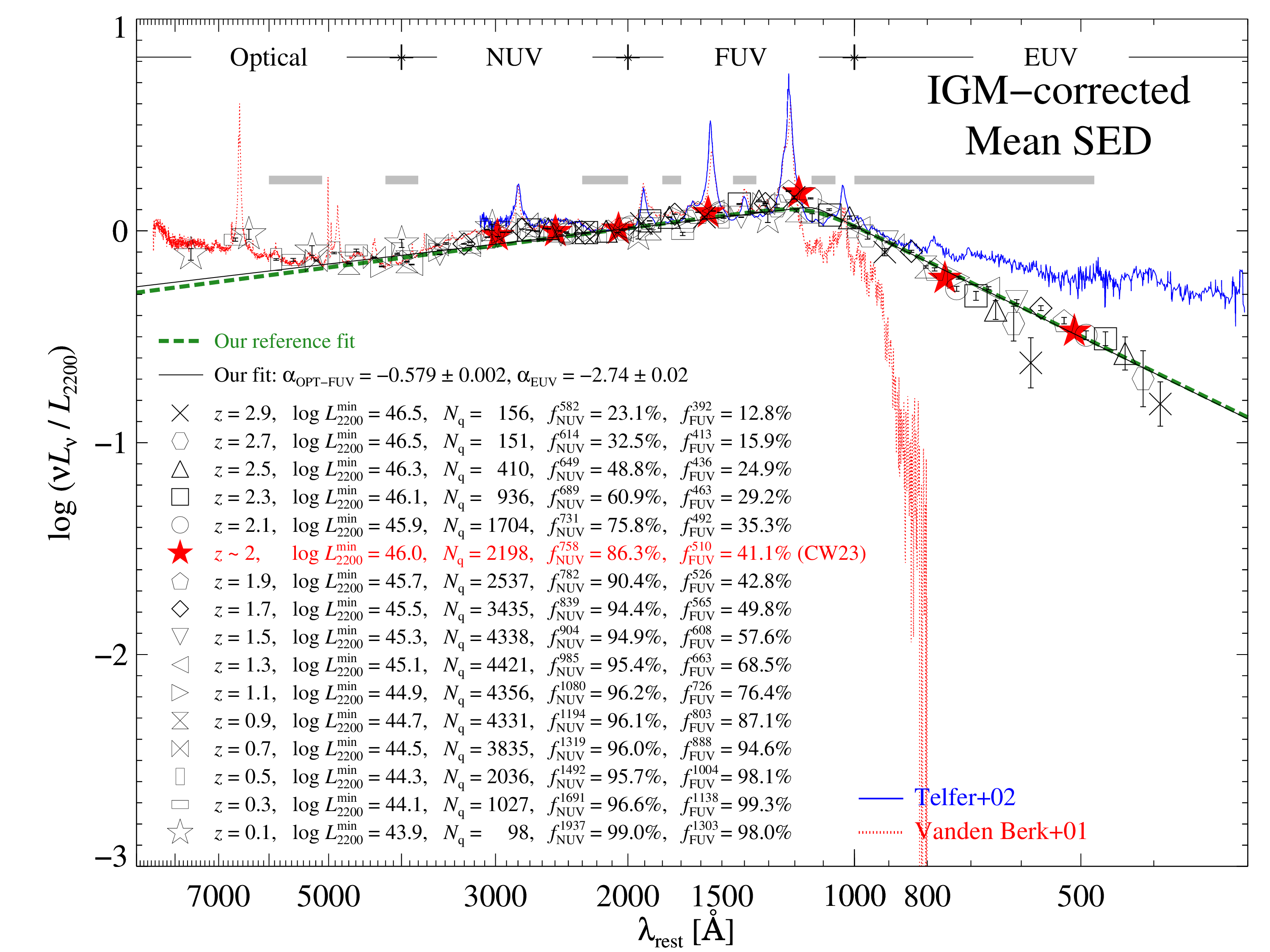}
    \includegraphics[width=0.42\linewidth]{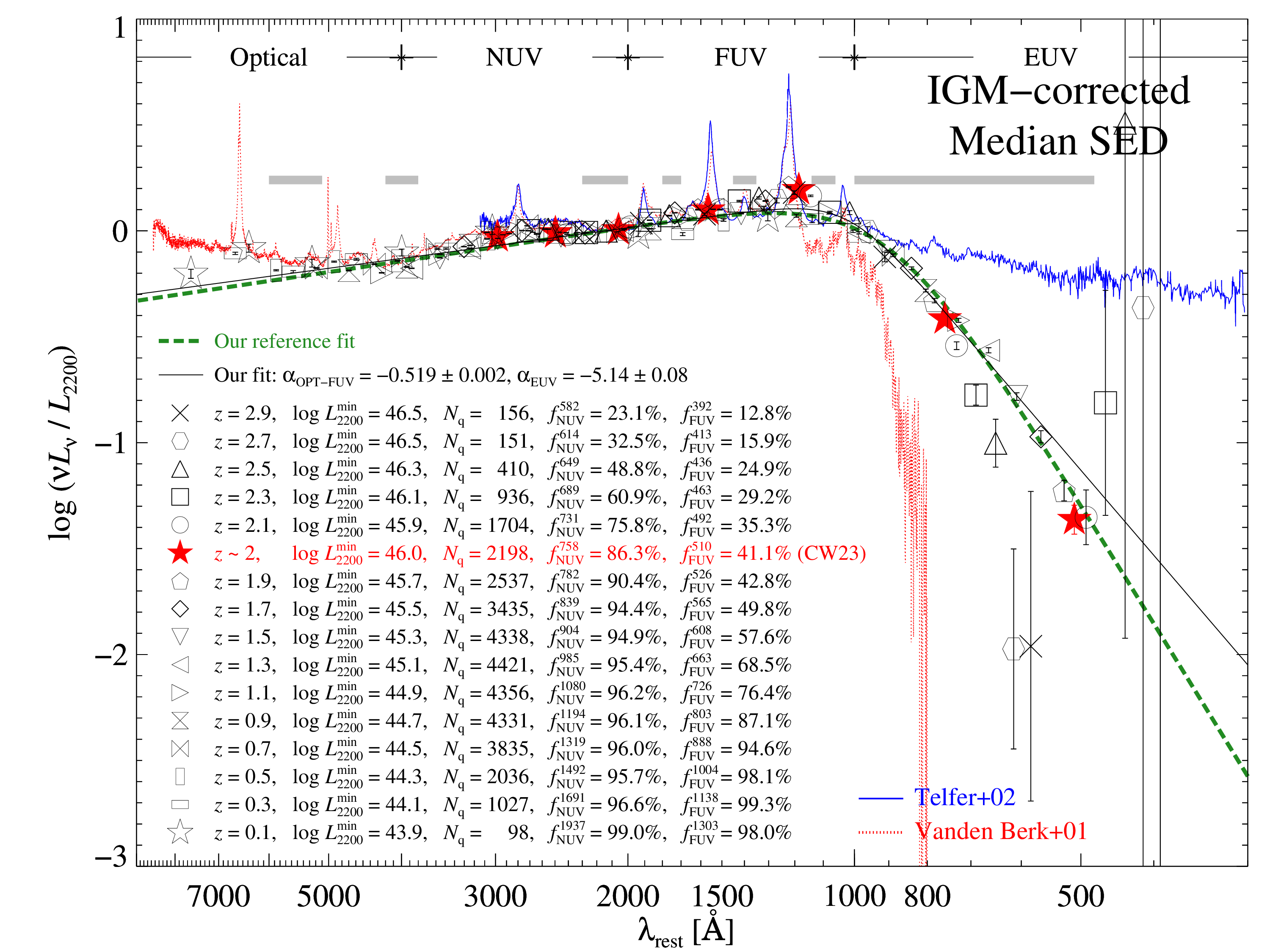}
\end{adjustwidth}
     \caption{{Panels} 
 in each row are the same as Figure \ref{fig:quasar_sed_composite}, but for the intrinsic quasar SEDs constructed from another two quasar samples selected with larger (top panels) or smaller (bottom panels) $\log L_{2200}^{\rm min}$, i.e., by $\pm 0.2$ dex in all redshift bins, than our reference values. In each panel, the green dashed curve is the best-fit smoothly broken power law taken from the corresponding panel of Figure \ref{fig:quasar_sed_composite}.}
     \label{fig:quasar_sed_composite_different_limits}
\end{figure}

{As introduced in Section \ref{sect:galex_detections}, we adopt redshift-dependent luminosity cutoffs to select the unique bright quasar sample, guided by the trade-off between high GALEX detection and large sample size (Figure \ref{fig:nq_ffuv_fnuv_fit}). One may consider other cutoffs, but it is hard to quantify which one is ``the best''. To convince the readers that the adopted cutoffs are indeed reasonable and that perturbations in some cutoffs will not lead to significantly different conclusions, we present in Figure \ref{fig:quasar_sed_composite_different_limits} the intrinsic mean/median quasar SEDs of another two quasar samples selected with larger or smaller $\log L_{2200}^{\rm min}$, i.e., by $\pm 0.2$ dex in all redshift bins, than our reference values (Section \ref{sect:galex_detections}). It is interesting to find that the universality of the intrinsic quasar SED preserves for changing the luminosity cutoffs across 0.4 dex. Although an even larger range for the luminosity cutoffs can be explored, the uncertainties induced by either a small sample size or low GALEX detection should be taken carefully.
}

\unskip

\subsection{How Can the Model-Predicted SEDs Be Properly Compared with the Observed Composite SEDs?}

In each redshift bin, there are many quasars with different physical properties, i.e., distinct $M_{\rm BH}$ and $\lambda_{\rm Edd}$. Previously, {we generated a single model-predicted SED for a quasar sample using its median value\endnote{Nominated in the legend of Figure 4 of \citet{Cai2023NatAs}, the median value of $\lambda_{\rm Edd}/M_{\rm BH}$ for the quasar sample in each luminosity bin was not correctly typed, although the resultant model-predicted SED is correct. Therein, the correct values for $\log(\lambda_{\rm Edd}/M_{\rm BH})^{\rm med}$ should be added by two.} of $\lambda_{\rm Edd}/M_{\rm BH}$ and compared} to the observed median quasar SED \cite{Cai2023NatAs}. But what about the mean SED? Here we construct the model-predicted mean/median composite SED in the same way as the observed mean/median composite one. Utilizing measurements on $M_{\rm BH}$ and $\lambda_{\rm Edd}$ from \citet{Rakshit2020ApJS..249...17R}, we generate a model-predicted SED for every quasar in a redshift bin and normalized all SEDs at $2200 \angstrom$ before obtaining the mean/median SED. 

Same as \citet{Cai2023NatAs}, we consider two typical disk models. One is the standard thin disk model \cite{Novikov1973blho.conf..343N}, whereas the other is the simply truncated disk model \cite{Laor2014MNRAS.438.3024L}, {in which the truncation radius is linked to a maximum temperature of $T_{\rm max} \simeq 3.6 \times 10^4 [(\lambda_{0.1}/M_9)/0.02]^{0.07} {\rm K}$, where $\lambda_{0.1} = \lambda_{\rm Edd} / 0.1$ and $M_9 = M_{\rm BH} / 10^9 M_\odot$. Please note that in both thin and truncated disk models adopted here, the model-predicted SED for a quasar is fixed as long as its $M_{\rm BH}$ and $\lambda_{\rm Edd}$ are given. In other words, there is no free parameter in these two models.}
Other models such as the slim disk \citep{Wang1999ApJ...522..839W}, the disk evaporation model \cite{Liu2009ApJ...707..233L,Qiao2013ApJ...777..102Q}, the condensation of the corona \cite{Liu2015ApJ...806..223L,Qiao2017MNRAS.467..898Q,Qiao2018MNRAS.477..210Q}, and the collapse of the disk \cite{Hagen2024MNRAS.tmp.2224H} are deferred to future works.

\begin{figure}[H]
    \includegraphics[width=0.475\linewidth]{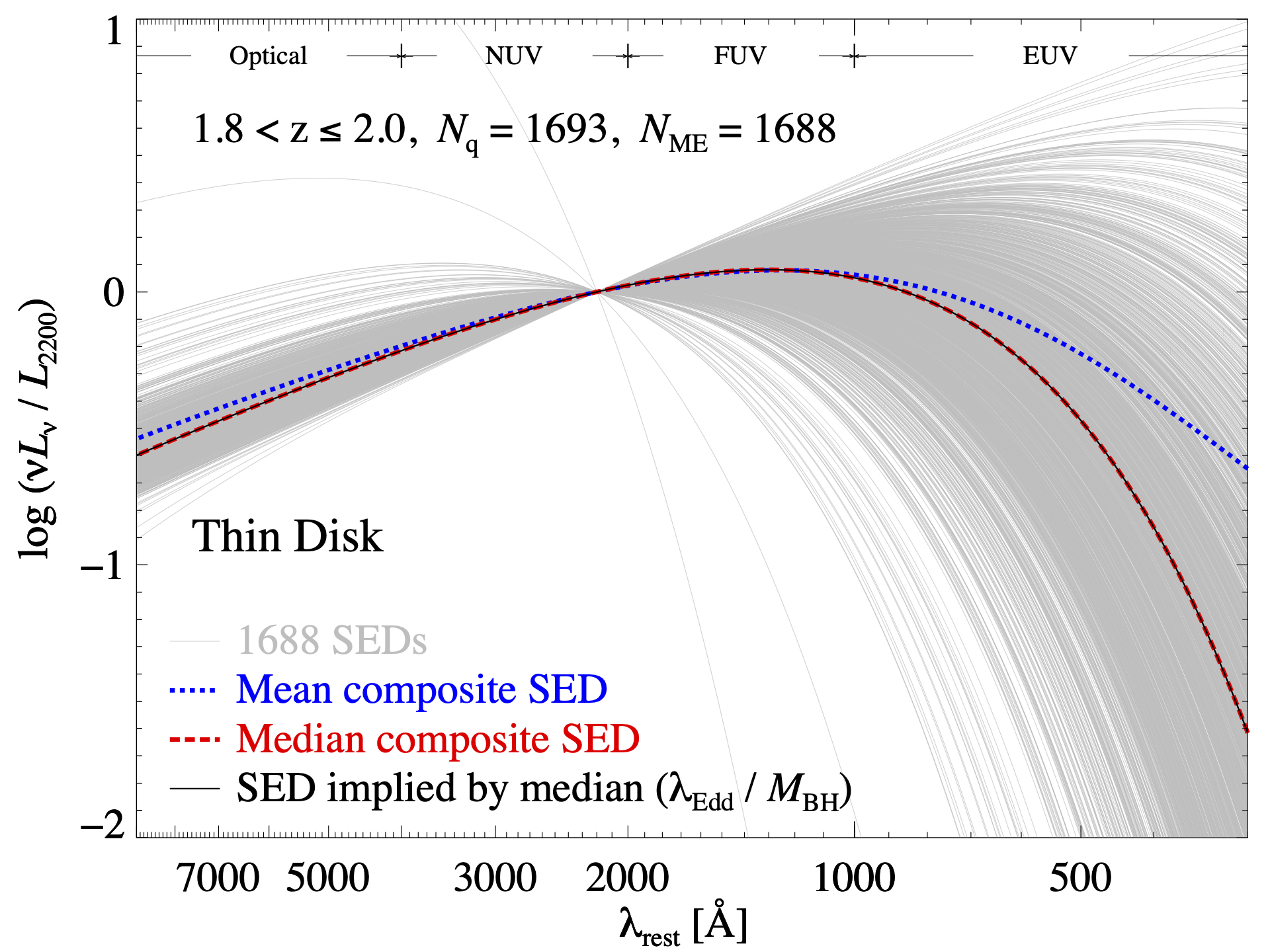}
    \includegraphics[width=0.475\linewidth]{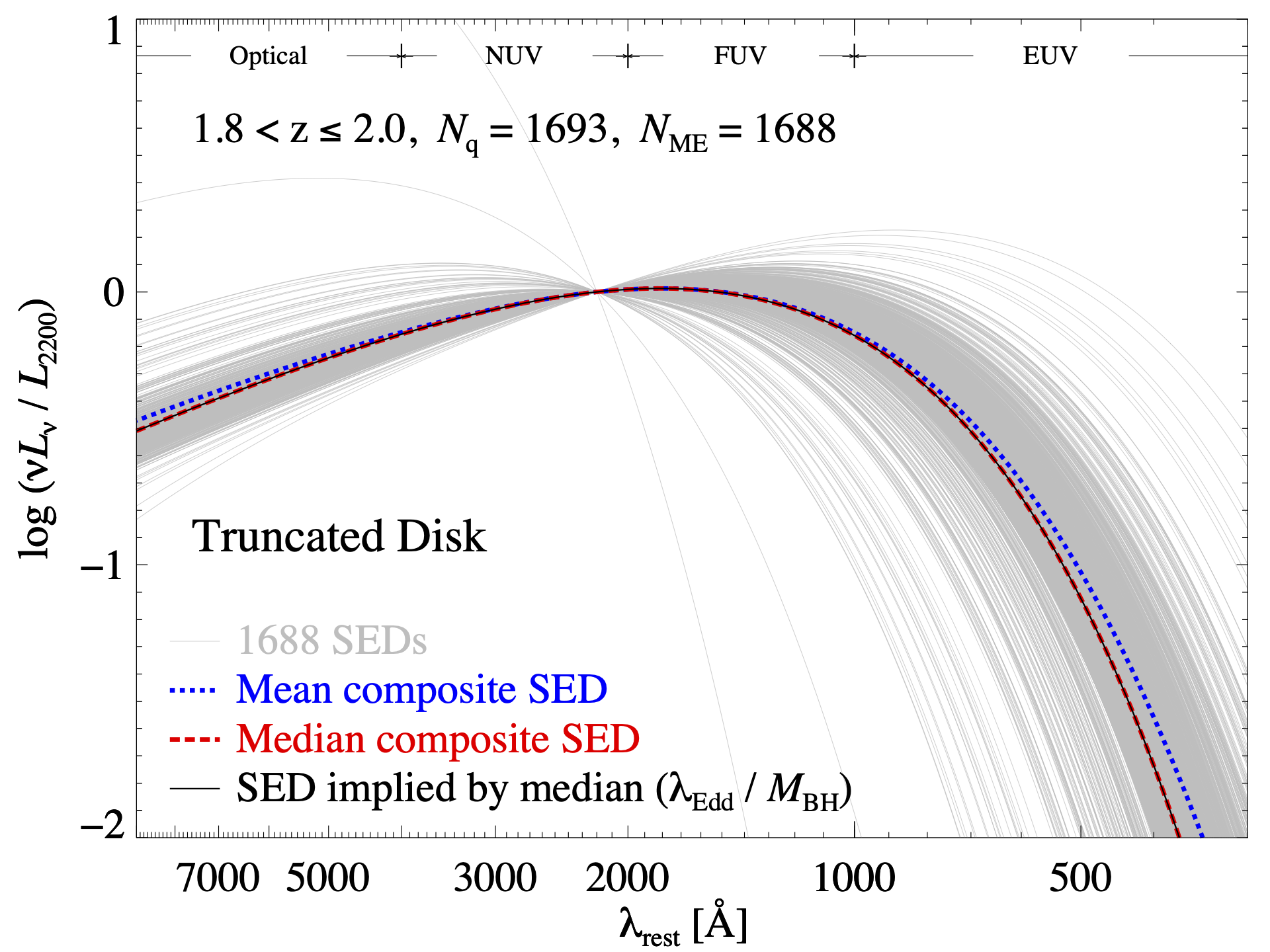}
     \caption{SEDs predicted by the thin disk (\textbf{left} panel) and truncated disk (\textbf{right} panel) models for $N_{\rm ME}$ quasars, with measurements on both $M_{\rm BH}$ and $\lambda_{\rm Edd}$, of $N_{\rm q}$ quasars in $1.8 < z \leqslant 2.0$ as an example. 
     Each gray thin solid curve is an SED of a quasar, while the blue dotted and red dashed curves are the resultant model-predicted mean and median composite SEDs for quasars in that redshift bin, respectively.
     For comparison, a black solid curve superimposed on the red dashed curve represents a specific SED implied by the median $\lambda_{\rm Edd} / M_{\rm BH}$ of the $N_{\rm ME}$ quasars.}
     \label{fig:sed_models}
\end{figure}

Figure \ref{fig:sed_models} shows SEDs predicted by both the thin and truncated disk models for quasars in $1.8 < z \leqslant 2.0$ as an example. For both disk models, we find that the SED implied by the median $\lambda_{\rm Edd}/M_{\rm BH}$ of a quasar sample is almost the same as the predicted median composite SED. The difference in the optical-to-FUV between the predicted mean and median composite SEDs is small. However, a large EUV difference between them is implied by the thin disk model but not by the truncated disk model. We confirm that these claims are also valid for quasar samples in other redshift bins. Consequently, directly comparing the model-predicted composite SED constructed in the same way as the observed composite SED is necessary, especially for the mean SED.

\unskip

\begin{figure}[H]
    \includegraphics[width=0.475\linewidth]{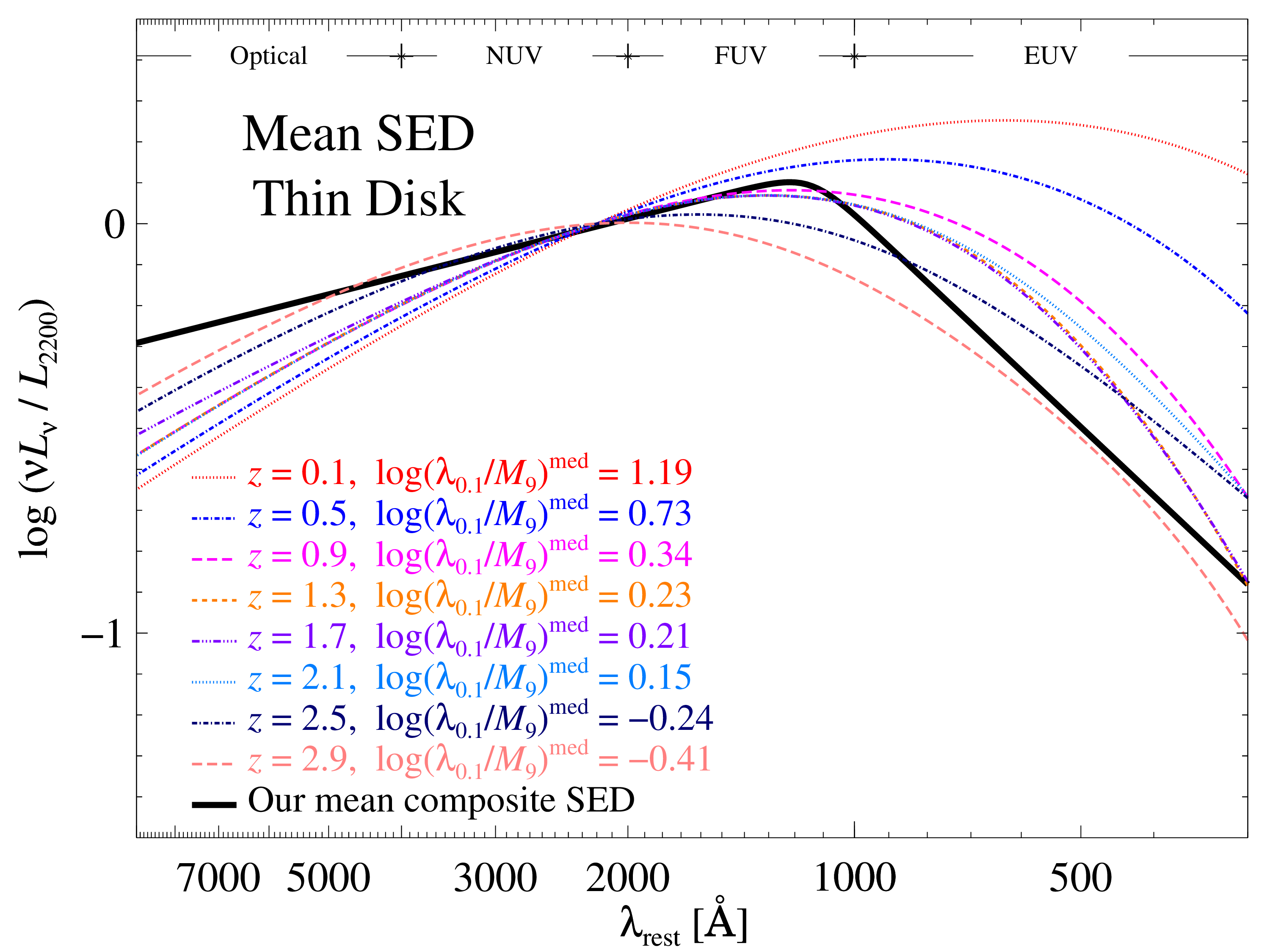}
    \includegraphics[width=0.475\linewidth]{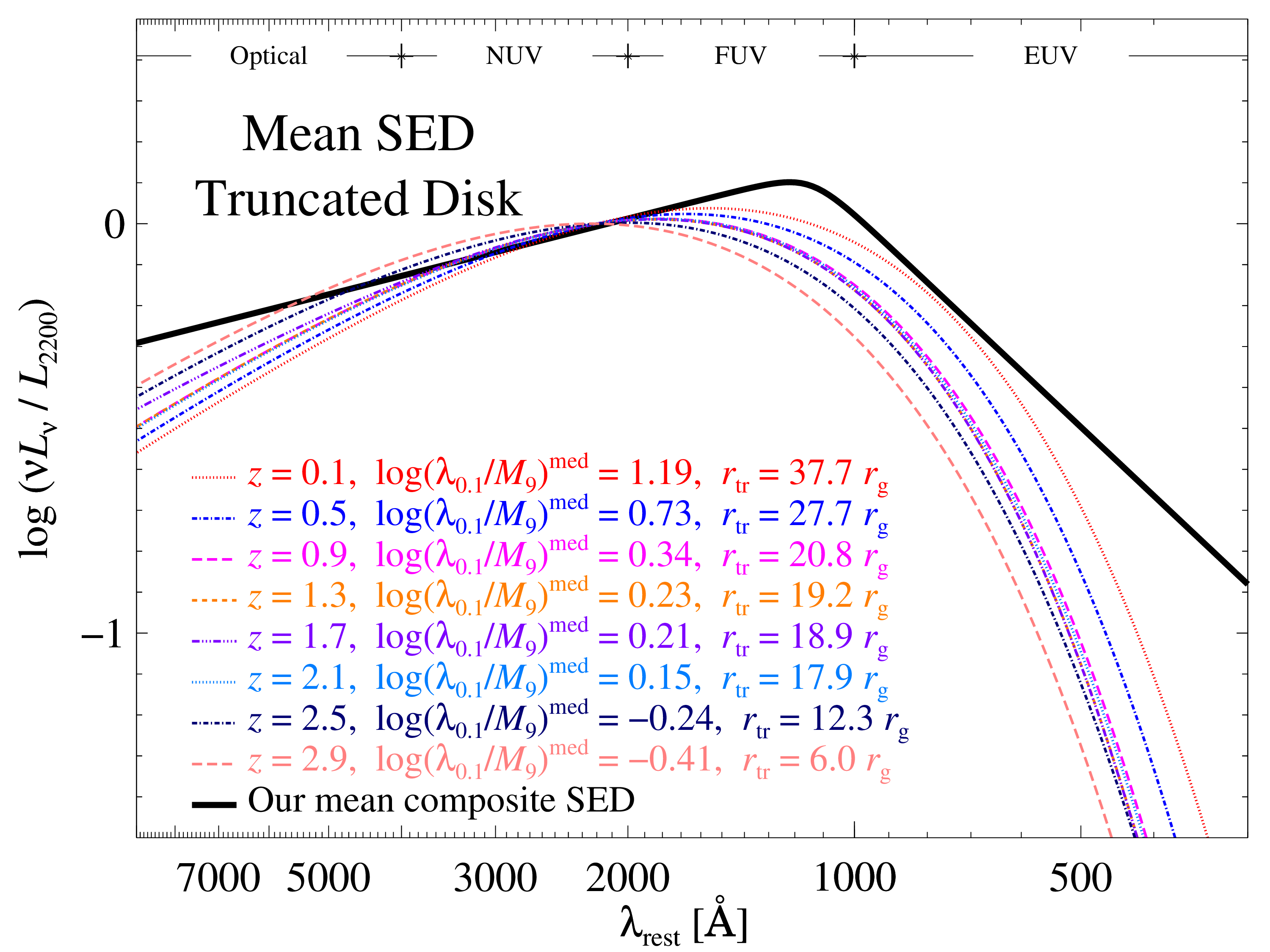}
    
    \includegraphics[width=0.475\linewidth]{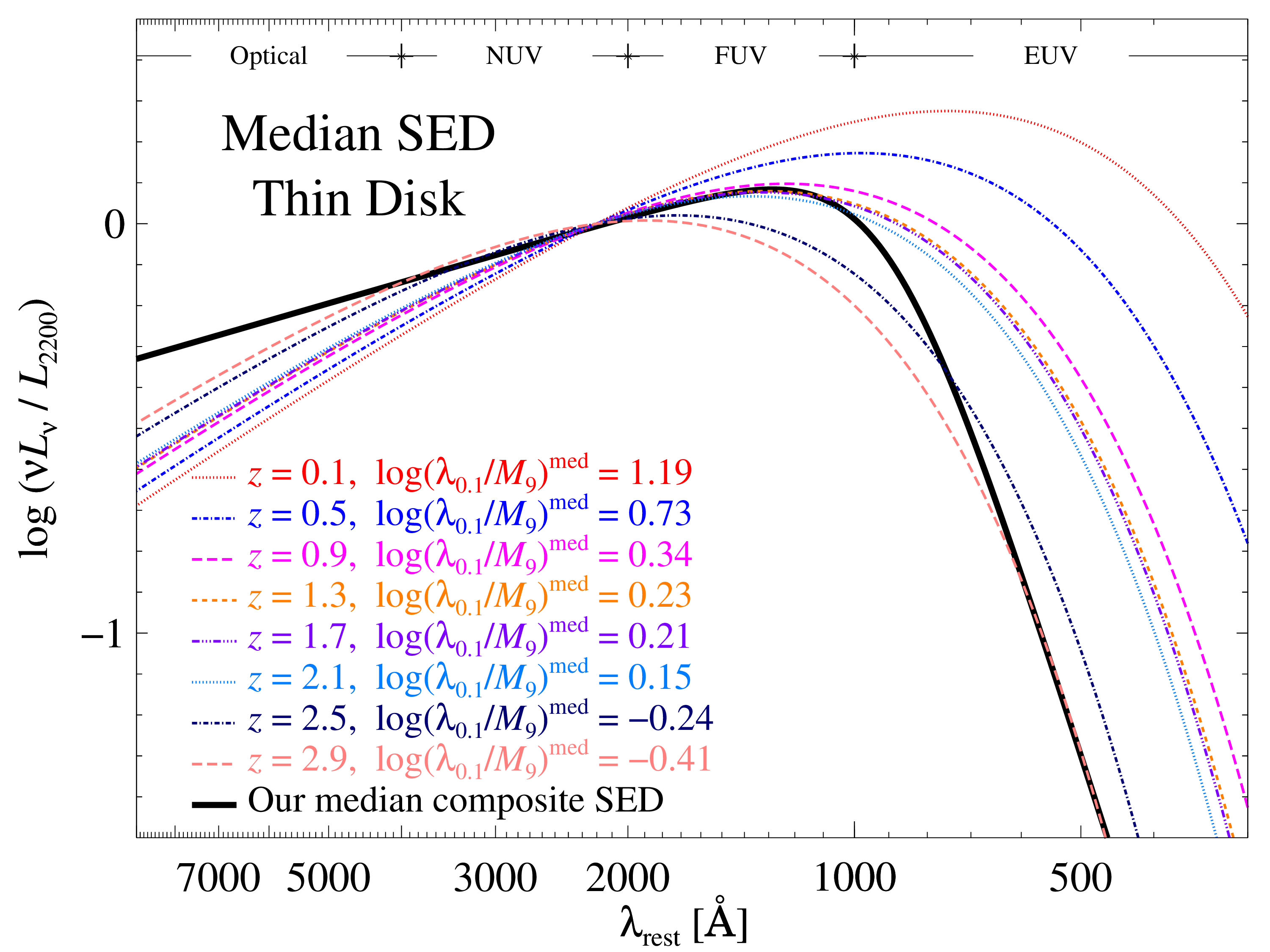}
    \includegraphics[width=0.475\linewidth]{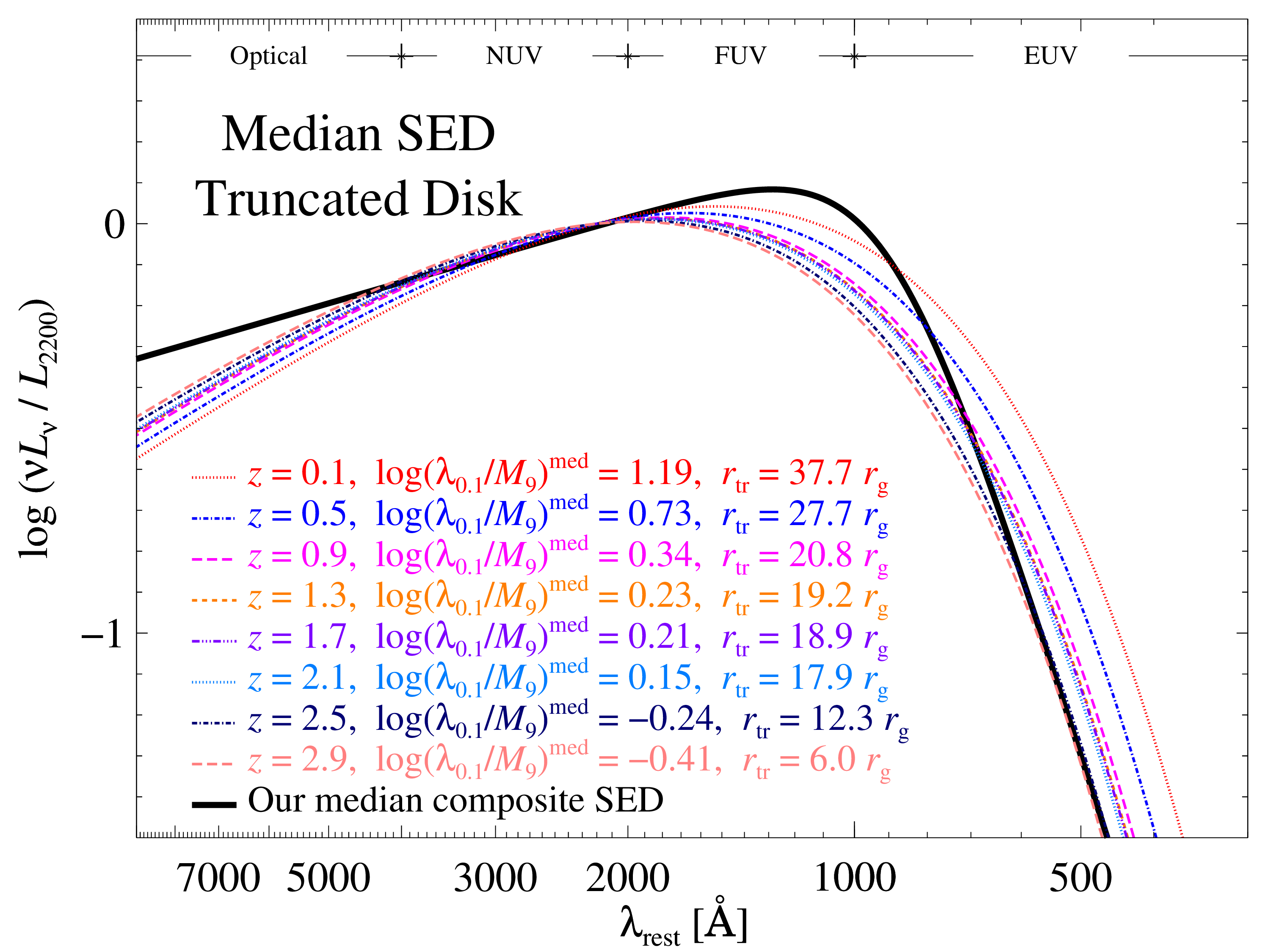}
     \caption{(The \textbf{top} panels): comparing the model-predicated mean composite SEDs for quasars in different redshift bins (colored thin curves) to the observed mean composite SED (black thick solid curve), the truncated disk model (\textbf{top-right} panel) performs better than the thin disk model (\textbf{top-left} panel). 
     The median values of $z$ and $\lambda_{\rm Edd}/M_{\rm BH}$ for quasars in each redshift bin are nominated in the legend. In the {right} panel for the truncated disk model, the legend contains a typical truncation radius, $r_{\rm tr}$ (in gravitational radius $r_{\rm g}$), corresponding to a maximum disk temperature determined by the median $\lambda_{\rm Edd}/M_{\rm BH}$ of quasars.
     (The \textbf{bottom} panels): same as the {top} ones, but for comparing the median composite SEDs.}
     \label{fig:sed_models_vs_fit}
\end{figure}
\unskip

\subsection{The Central Engine at Work for Quasars Remains an Enigma}

Although the much steeper EUV decrease of the median composite SED compared to the mean SED predicted by the thin disk model for quasars in each redshift bin (the left panel of Figure \ref{fig:sed_models}) is similar to the observed difference between the mean and median composite quasar SEDs (Figure \ref{fig:quasar_sed_composite}), the left panels of Figure \ref{fig:sed_models_vs_fit} demonstrate that both the mean and median composite SEDs predicted by the thin disk model are strongly dependent on redshift and thus in conflict with the redshift independence of both the mean and median composite quasar SEDs observed. 

Instead, the mean and median composite SEDs predicted by the truncated disk model are comparable among redshift bins (the right panel of Figure \ref{fig:sed_models}) and are only somewhat dependent on redshift (the right panels of Figure \ref{fig:sed_models_vs_fit}), much weaker than predictions of the thin disk model. 
{In the truncated disk model, it is the similar maximum temperature, which is attributed to a microscopic atomic origin and thus is weakly dependent on both $M_{\rm BH}$ and $\lambda_{\rm Edd}$ \citep{Laor2014MNRAS.438.3024L}, such that the truncated disk model can give rise to relatively universal SEDs for quasars at different redshifts, where their typical $M_{\rm BH}$ and $\lambda_{\rm Edd}$ are distinct (Figure \ref{fig:physical_properties_z}). }
For quasars at different redshifts, the median composite SEDs predicted by the truncated disk model are very close to the observed median composite quasar SED, though there are some discrepancies (the bottom-right panel of Figure \ref{fig:sed_models_vs_fit}). More significant discrepancies can be found in comparing the mean composite SEDs where the truncated disk model predicts EUV-redder mean SEDs with increasing redshift (the bottom-left panel of Figure \ref{fig:sed_models_vs_fit}). 

Since the physical properties of quasars are significantly dependent on redshift (Figure \ref{fig:physical_properties_z}), the universal mean/median composite quasar SED could not be the result of the classic thin disk model (the left panels of Figure \ref{fig:sed_models_vs_fit}). Instead, the truncated disk model provides a possibility for the quasar central engine, though clear discrepancies still exist (the right panels of Figure \ref{fig:sed_models_vs_fit}). 
Please note that the two models considered here cannot account for the corona emission and soft X-ray excess {as well as the ubiquitous AGN variability}. 

{We note that} models including X-ray emission {may not} be easily responsible for the universality of our composite quasar SED, which is independent of many physical properties of quasars. {For example, in the X-ray model proposed by \mbox{\citet{Kubota2018MNRAS.480.1247K}},  the warm corona mainly contributes the EUV emission, and parameters relevant to the warm corona are all free regardless of $M_{\rm BH}$ and $\lambda_{\rm Edd}$. A tentative fit using this model indeed finds that achieving model-predicted SEDs comparable to our universal composite quasar SED is possible for different $M_{\rm BH}$ and $\lambda_{\rm Edd}$ but in a fine-tuning way, and that the derived dependences of parameters on $M_{\rm BH}$ and $\lambda_{\rm Edd}$ are obscure (J. L. Kang et al. 2025, in preparation). 
Another interesting X-ray model to be considered is the disk evaporation model by (\cite{Taam2012ApJ...759...65T}, see their Equation (6)) where the truncation radius does not depend on $M_{\rm BH}$ and is weakly dependent on the viscosity parameter, but strongly depends on $\lambda_{\rm Edd}$ and the ratio of gas pressure to the total pressure.}

Extending our SED analyses from EUV to (soft/hard) X-ray is essential for understanding the quasar central engine. Unfortunately, the X-ray detections of the SDSS-selected quasars are exceedingly low so far. For example, even for our unique bright quasar sample (Figure \ref{fig:quasar_selection}), only $\simeq$$7\%$ have soft/hard X-ray measurements by {\it {XMM-Newton}} according to the multi-wavelength cross-matched catalog of \citet{Paris2018A&A...613A..51P}. {Since the {\it XMM-Newton} observations do not cover most SDSS quasars, the real {\it XMM-Newton} detection fraction should be larger than $7\%$ but is hard to assess. The situation in the soft X-ray is much better thanks to the extended ROentgen Survey with an Imaging Telescope Array (eROSITA) instrument aboard the Spectrum-Roentgen-Gamma (SRG) mission \cite{Predehl2021A&A...647A...1P}.} Considering $\sim$$9600$ quasars in our unique bright quasar sample within the German footprints of the first six months of survey operations (eRASS1; \cite{Merloni2024A&A...682A..34M}) conducted by eROSITA, $\simeq$$38\%$ have counterparts in the eRASS1 main catalog within a matching radius of 15 arcsec (S. J. Chen et al., 2025, in preparation). Excitingly, a continuing increase in the eROSITA detection can be expected once subsequent deeper eROSITA data releases are available.
At the moment, we caution again that extending the SED analyses to X-ray by simply dropping the X-ray non-detections may easily result in improper conclusions.

\subsection{Properties of Dust and Gas in the Quasar Host Galaxies}

When constructing the intrinsic mean/median SEDs for quasars at different redshifts, the dust attenuation has been neglected. More uncertain than the amount of dust in the quasar host galaxies is the EUV shape of the dust-attenuation curve. Actually, no measurement on the dust-attenuation curve has been available at wavelengths shorter than the rest-frame $\sim$$1000 \angstrom$ until recently. \citet{Gaskell2024MNRAS.533.3676G} derived, for the first time, a mean attenuation curve down to the rest-frame $\sim$$400 \angstrom$ by (1) comparing the SED difference between normal quasars and quasars with broad absorption lines (BAL) and (2) associating the SED difference with an extra dust absorption along the LOS to BAL quasars. 
The derived mean attenuation curve in the optical-to-FUV is very similar to the attenuation curve of the Small Magellanic Cloud (SMC; \cite{Weingartner2001ApJ...548..296W}) and not to the relatively flat mean attenuation curve of normal quasars \cite{Gaskell2007arXiv0711.1013G}, suggesting that the dust in the host galaxy along the LOS to normal quasars has distinct properties from the dust likely in clouds outflowing from the broad line region (BLR) along the LOS to BAL quasars.

Although the mean attenuation curve derived by \citet{Gaskell2024MNRAS.533.3676G} has a large dispersion in the EUV, it suggests a more significant attenuation over the whole EUV and an attenuation peak at a wavelength shorter than predicted by \mbox{\citet{Weingartner2001ApJ...548..296W}}. If the mean attenuation curve for our quasars has a peak in the EUV, likely around $\sim$700--500 $\angstrom$ as predicted by \citet{Weingartner2001ApJ...548..296W} and measured by \citet{Gaskell2024MNRAS.533.3676G}, there would be a resultant dip (or concavity) in the composite mean/median quasar SED, which is however not observed.
Therefore, the universality and the smoothness of our mean/median composite quasar SED suggests that (1) the amount of dust is little in the LOS of most normal quasars, such as our quasars brighter than $\log L_{\rm bol} \simeq 45.5$ (see also Figure 5 of \citet{Gaskell2004ApJ...616..147G}), and/or (2) the AGN attenuation curve is very flat extending from the optical-to-FUV \cite{Gaskell2007arXiv0711.1013G} to the EUV for normal quasars. 

The EUV ionizing continuum is also very sensitive to the amount of neutral hydrogen. Similarly, the universality of our mean/median composite quasar SED suggests that the interstellar and circumgalactic mediums along the LOS to quasars should be gas-free (or highly ionized). This is possible since bright quasars are more frequently found in massive early-type galaxies, which generally have a low gas content. And it is enough that the LOS to a quasar is gas-free, rather than the whole host galaxy of the quasar, given the general anisotropic radiation of the accretion disk \cite{Kato2008bhad.book.....K} and the existence of the dusty torus and the ionization cone \cite{Wu2023MNRAS.522.1108W}. When the LOS of a quasar host galaxy has an abundant interstellar medium, the ``quasar'' is obscured and may be undetected or too faint to be identified as a quasar.

If there were a redshift dependence of the dust attenuation and/or host-related hydrogen absorption, a fine-tuning of the dust attenuation and/or hydrogen absorption as a function of redshift would be unavoidable. Otherwise, it is hard to understand why the observed quasar SEDs at different redshifts are nicely consistent with each other after dust attenuation and/or hydrogen absorption. Instead, the universality of our mean/median composite quasar SED may imply that properties of dust attenuation and/or hydrogen absorption, if any, in the quasar host galaxies at $z \lesssim 3$ should be similar on average, independent of redshift and luminosity. In future, it would be interesting to, after taking the dust attenuation and/or hydrogen absorption into account, directly compare the model-predicted distribution of $\log(L_w/L_{2200})$ in the EUV to the observed asymmetric one, indicated by the huge difference between the mean and median composite quasar SEDs in the EUV (Figure \ref{fig:quasar_sed_composite}).

\subsection{\bf Evolutionary of Quasars Since Cosmic Noon}

{The decline in the maximum luminosity of quasars since $z$$\sim$$3$ (Figure \ref{fig:quasar_selection}) intuitively suggests an evolutionary where the most violent BH accretion shifts from quasars at $z$$\sim$2--3 with $L_{\rm bol}$$\sim$$10^{47} {\rm erg s^{-1}}$, $M_{\rm BH}$$\sim$$10^{9.5} M_\odot$ and $\lambda_{\rm Edd}$$\sim$$0.3$ to those at $z$$\sim$$0.1$ with $L_{\rm bol}$$\sim$$10^{45} {\rm erg s^{-1}}$, $M_{\rm BH}$$\sim$$10^{8} M_\odot$, and $\lambda$$\sim$$0.1$ (Figure \ref{fig:physical_properties_z}). Why are there few low-redshift quasars as bright as those at cosmic noon? Our universal quasar SED yields new constraints on the evolution of quasars. As hinted by the simple truncated disk model, the accretion disk of a quasar with a less massive BH likely has a larger truncation radius (the right panels of Figure \ref{fig:sed_models_vs_fit}). If the disk wind indeed plays a role in determining the truncation radius, it would also result in a lower radiative efficiency for the accretion to a less massive BH \citep{Laor2014MNRAS.438.3024L}, and in the meantime, slow down the BH growth. Therefore, when quantitatively comparing the model to data, including the cosmological evolution of the quasar luminosity function and the BH mass function, it is worthy of consideration of a more realistic accretion model with a mass-dependent radiative efficiency rather than a simple model with a conventional constant radiative efficiency of $\simeq$$0.1$.
}

\subsection{Implications of a Deficit of the EUV Ionizing Radiation}

As illustrated in Figure \ref{fig:quasar_sed_composite}, our mean composite quasar SED at the rest-frame wavelengths beyond $\sim$$800 \angstrom$ is nicely consistent with but redder at wavelengths short of $\sim$$800 \angstrom$ than, the mean composite spectrum of \citet{Telfer2002ApJ}, which is the EUV-reddest one among previous mean composite quasar spectra \cite{Scott2004ApJ,Stevans2014ApJ,Lusso2015MNRAS}. Our median composite quasar SED is even redder in the EUV than our mean composite quasar SED. Both our mean and median composite quasar SEDs indicate a deficit of the EUV ionizing continuum. Relative to the mean composite spectrum of \citet{Telfer2002ApJ}, the amount of EUV ionizing continuum, i.e., the integral from $912 \angstrom$ to $300 \angstrom$, implied by our best-fit smoothly broken power law for the mean (median) composite quasar SED is smaller by a factor of $\simeq$$1.6$ ($\simeq$$3.6$). Therefore, the deficit of the EUV ionizing continuum has important implications for, e.g., the production of broad emission lines and cosmic reionization.

\subsubsection{Production of Broad Emission Lines}

Broad emission lines, such as $\Lya$, $\CIVoffn$, $\MgIItsne$, $\Hb$, and $\Ha$, are prominent characteristics of quasars, and are generally discussed within three cloud models: magnetic pressure conﬁned clouds \cite{Rees1989ApJ...347..640R}, locally optimally emitting clouds \cite{Baldwin1995ApJ...455L.119B}, and radiation pressure-confined clouds \cite{Dopita2002ApJ...572..753D}. Regardless of the different physics and conditions involved in these cloud models and the fact that, hitherto, no such models have reproduced all known line properties \cite{Netzer2020MNRAS.494.1611N,Ferland2020MNRAS.494.5917F}, photoionization owing to the incident ionizing radiation of the central quasar is a key ingredient for generating these emission lines. 

\begin{figure}[H]
  \hspace{-2em}  \includegraphics[width=0.8\linewidth]{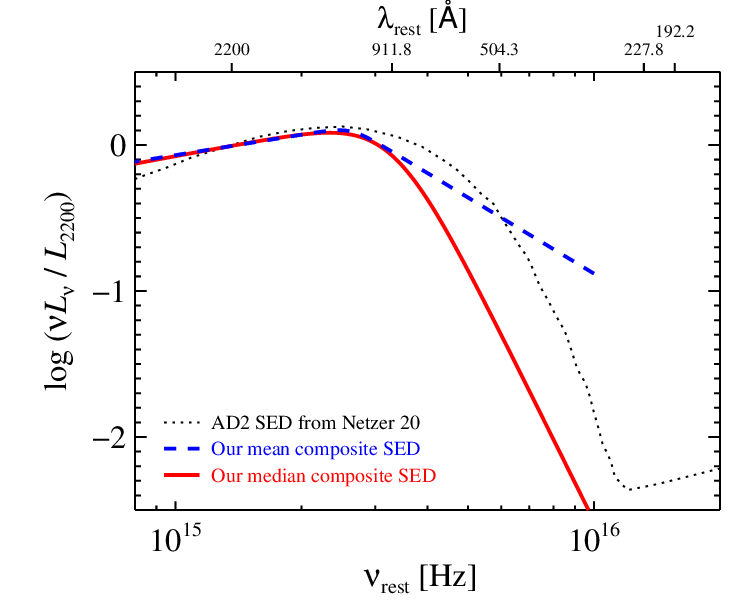}
     \caption{The blue dashed (red solid) curve is our mean (median) composite quasar SED. The black dotted curve is the AD2 SED of \citet{Netzer2020MNRAS.494.1611N}, whose ionizing continuum is weakest among the four SEDs adopted by \citet{Netzer2020MNRAS.494.1611N} and has difficulties in producing large enough line luminosities, including \CIVoffn, \HeIIosfz, and \Hb. On the top axis, there are marked specific wavelengths corresponding to the ionization potential energies of \HI ($911.8 \angstrom$), \HeI ($504.3 \angstrom$), \HeII ($227.8 \angstrom$), and \CIV ($192.2 \angstrom$).}
     \label{fig:netzer20_fig1_seds_AD2}
\end{figure}

Less ionizing radiation usually results in weaker line emission. Combining the radiation pressure-confined clouds with a theoretical SED that has relatively weak EUV emission (cf. the AD2 SED in Figure 1 of \citet{Netzer2020MNRAS.494.1611N} and the black dotted curve in Figure \ref{fig:netzer20_fig1_seds_AD2}), \citet{Netzer2020MNRAS.494.1611N} showed in his Figure 3 that the AD2 SED fails to produce large enough line luminosities, including \CIVoffn, \HeIIosfz, and \Hb. As illustrated in Figure \ref{fig:netzer20_fig1_seds_AD2}, our mean and median composite quasar SEDs have even weaker EUV emissions than the AD2 SED and thus would give rise to smaller line luminosities according to the model of \citet{Netzer2020MNRAS.494.1611N}. Our median composite quasar SED implies that more than 50\% of quasars with broad emission lines have a strong deficit of the EUV emission.

Given the very low EUV emission, why the emission lines are there is an interesting question to be addressed. Although detailed calculations on the line production are demanded, our EUV-deficit composite SEDs hint at that (1) the cloud density may be smaller and/or (2) the role of collision excitation or the other energy sources, such as the magnetic energy associated with the magnetic pressure-confined clouds, may be more important than expected. 
Another possibility for the large EUV deficit found in the median composite quasar SED is that the EUV emission seen by the BLR clouds is not attenuated by dust within the BLR and thus is different from that observed by us, which is attenuated by dust in the quasar host galaxy. On the one hand, examining differences in the broad emission lines among sub-samples of our unique bright quasars divided by $L_w/L_{2200}$, where $w$ is shorter than $\sim$$1000 \angstrom$, can tell us whether the EUV emission seen by the BLR clouds is attenuated: no or little difference indicates unattenuated EUV emission for clouds. On the other hand, examining the relationship between $L_w/L_{2200}$ and $L_{\rm d}/L_{2200}$, where $L_{\rm d}$ is the dust luminosity at wavelengths likely spanning from near- to far-infrared, could suggest whether the EUV emission is attenuated by dust in the quasar host galaxy: a negative correlation between them may indicate dust attenuation up to the host scale.

\subsubsection{Cosmic Reionization}

Two important episodes in cosmic history are the reionization of neutral hydrogen and helium. The relatively well-understood helium reionization completing at $z$$\sim$$3$ is mainly contributed by quasars \citep{Haardt2012ApJ...746..125H}, whereas the exact process of hydrogen reionization, though likely ending around $z $$\sim$5--6, and the major ionizing contributor, e.g., star-forming galaxies \cite{Robertson2010Natur.468...49R,Cai2014ApJ...785...65C,Jiang2016ApJ...833..222J,Onoue2017ApJ...847L..15O,Matsuoka2018ApJ...869..150M,Parsa2018MNRAS.474.2904P,Kulkarni2019MNRAS.488.1035K,Jiang2022NatAs...6..850J,Atek2024Natur.626..975A} or quasars/AGN \cite{Volonteri2009ApJ...703.2113V,Giallongo2015A&A...578A..83G,Madau2015ApJ...813L...8M,Giallongo2019ApJ...884...19G,Grazian2022ApJ...924...62G,Madau2024ApJ...971...75M}, are still under debate, especially after the discovery of a large population of moderate-luminosity type 1 AGN at $4 \lesssim z \lesssim 13$ by the James Webb Space Telescope \cite{Kocevski2023ApJ...954L...4K}. 

The cosmic reionization is more relevant to the mean ionizing radiation and, thus, to the mean composite quasar SED. Here, our finding of an EUV-deficit mean quasar SED may imply a smaller AGN contribution to the hydrogen reionization. Note that properties of dust and gas in the host galaxies of quasars at $z > 5$ are not necessarily similar to those of the quasar host galaxies at $z \lesssim 3$ (see \citet{Brooks2024arXiv241007340B} for a potential sample of heavily dust-attenuated AGN at $z \gtrsim 3.5$ owing to dust in the BLR but not in the host galaxy) and neither does the mean composite quasar SED. Directly extending our SED analysis to quasars at higher redshifts with the help of future deep ultraviolet facilities would provide important constraints on the mean escape fraction of the Lyman continuum radiation of quasars. Moreover, more consequences of our EUV-deficit mean composite quasar SED on the helium reionization, the cosmic UV background, and the thermal history of the IGM are worthy of investigation.

\section{Conclusions}
\label{sect:conclusion}

We consider a unique bright quasar sample with 23,256 sources selected from the SDSS DR14Q catalog and covered by the GALEX tiles with FUV exposure times longer than 1000 s and FOV offsets smaller than $0.5^\circ$. Combining the SDSS and GALEX photometry, bias-free mean and median SEDs of quasars in 15 redshift bins, i.e., from $z = 0$ to $z = 3$ in bins of 0.2, are obtained after taking the GALEX non-detection into account. These bias-free SEDs of quasars at different redshifts are very similar and are consistent with previous mean composite quasar spectra at wavelengths beyond $\simeq$$1000 \angstrom$. Instead, at wavelengths shorter than $\simeq$$1000 \angstrom$, these bias-free SEDs become redder in the EUV with increasing redshift, in line with the increasing significance of the IGM absorption. 

After correcting for the IGM absorption, these intrinsic mean/median SEDs of quasars at different redshifts are surprisingly similar, i.e., independent of redshift, and form a universal mean/median composite SED for quasars. Our mean composite quasar SED is redder in the EUV than the previous EUV-reddest mean composite quasar spectrum, whereas our median composite quasar SED is even redder in the EUV. Furthermore, our mean/median composite quasar SED is plausibly independent of both the BH mass and the Eddington ratio, which clearly depends on redshift.

Considering two typical disk models, i.e., the standard thin disk model and the simply truncated disk model, we demonstrate that directly comparing the model-predicted composite SED constructed in the same way as the observed composite SED is necessary, especially for the mean one. Thus, we show that our universal mean/median composite quasar SED prefers the simply truncated disk model rather than the standard thin disk model, though there are discrepancies requiring more sophisticated models.

Regardless of the uncertain physics of the quasar central engine, the universality of our mean/median composite quasar SED strongly suggests that properties of dust and gas in the quasar host galaxies at $z \lesssim 3$ are similar. Otherwise, fine-tuning for both the dust content and the gas ionization state as a function of redshift is unavoidable. Furthermore, given our EUV-deficit mean/median composite quasar SED, implications for the production of broad emission lines of quasars and the cosmic reionization by quasars are discussed.

Complemented by the all-sky X-ray surveys, such as eROSITA, and with the help of future deep ultraviolet facilities, such as the China Space Station Telescope \cite{Gong2019ApJ...883..203G} and the Ultraviolet Explorer \cite{Kulkarni2021arXiv211115608K}, revolutions in our understanding of the quasar central engine {as well as many related aspects are promising. Therefore, exploring the discovery capability of these UV and X-ray facilities deserves further comprehensive work.}

\vspace{6pt}


\funding{{This} 
 work is funded by National Key R\&D Program of China (grant No. 2023YFA1607903) and the National Science Foundation of China (grant Nos. 12373016 and 12033006). Z.C. acknowledges support from the China Manned Space Project under grant No. CMS-CSST-2021-A06 and the Cyrus Chun Ying Tang Foundations.
}

\dataavailability{{Data for the quasar samples, the filter-weighted broadband IGM transmissions, and the universal intrinsic mean/median composite quasar SED are available {at} 
 \url{http://staff.ustc.edu.cn/\~zcai/agnsed_c24/}, while} the other data underlying this article will be shared upon request to the corresponding author.}

\acknowledgments{{I}  
 am grateful {to the three anonymous referees for their constructive suggestions}, and to Gianfranco De Zotti, {Erlin Qiao}, Mouyuan Sun, and Jun-Xian Wang for valuable comments. I also thank Shi-Jiang Chen for examining the eROSITA counterparts to our quasar sample, {and Jia-Lai Kang for fitting other disk model to our universal composite quasar SED}. 
On the occasion of celebrating the 10th Anniversary of the Re-establishment of the Department of Astronomy at Xiamen University (XMU), I sincerely appreciate the admission to astronomy afforded by Wei-Min Gu and the joint PhD project between XMU and SISSA fostered by Ju-Fu Lu.
}

\conflictsofinterest{The author declares no conflicts of interest.}
\newpage
\abbreviations{Abbreviations}{
The following abbreviations are used in this manuscript:\\

\noindent 
\begin{tabular}{@{}ll}
AGN & Active Galactic Nucleus/Nuclei\\
BAL & Broad Absorption Line\\
BH & Black Hole\\
BLR & Broad Line Region\\
DR14Q & Data Release 14 Quasar\\
eROSITA & extended ROentgen Survey with an Imaging Telescope Array\\
EUV & Extreme Ultraviolet\\
FOV & Field of View\\
FUV & Far Ultraviolet\\
GALEX & Galaxy Evolution Explorer\\
IGM & Intergalactic Medium\\
LOS & Line of Sight\\
NUV & Near-ultraviolet\\
SDSS & Sloan Digital Sky Survey\\
SED & Spectral Energy Distribution\\
SMC & Small Magellanic Cloud \\
SRG & Spectrum-Roentgen-Gamma\\
UV & Ultraviolet
\end{tabular}
}


\begin{adjustwidth}{-\extralength}{0cm}
\printendnotes[custom]

\reftitle{References}


\PublishersNote{}
\end{adjustwidth}
\end{document}